\documentclass{bull-l}

\usepackage{graphicx}
\usepackage{color}
\usepackage{amssymb}
\usepackage{amsfonts}
\usepackage{amsmath}
\usepackage{csquotes}
\usepackage{hyperref}
\usepackage{marginnote}

\theoremstyle{definition}

\theoremstyle{remark}

\numberwithin{equation}{section}
\def\cL{\mathcal{L}}
\def\cM{\mathcal{M}}
\def\cT{\mathcal{T}}
\def\Real{\mathbb{ R}}
\newcommand{\be}{\begin{equation}}
\newcommand{\ee}{\end{equation}}

\begin{document}

\title[The Emergence of Gravitational Wave Science]
      {The Emergence of Gravitational Wave Science: \\
       100 Years of Development of \\
       Mathematical Theory, 
       Detectors, \\
       Numerical Algorithms, 
       and
       Data Analysis Tools 
       }


\def\addUMICH{Instituto de F\'{\i}sica y Matem\'aticas, Universidad Michoacana de San Nicol\'as de Hidalgo,
Edificio C-3, Ciudad Universitaria, 58040 Morelia, Michoac\'an, M\'exico,}

\def\addUCSDa{Department of Mathematics, Department of Physics, University of California San Diego, La Jolla California 92093, USA}

\def\addUCSDb{Center for Computational Mathematics, Center for Astrophysics and Space Sciences, University of California San Diego, La Jolla California 92093, USA}

\def\addUCSDc{San Diego Supercomputer Center, University of California San Diego, La Jolla California 92093, USA.}

\def\addJPL{Jet Propulsion Laboratory, California Institute of Technology, 4800 Oak Grove Drive, Pasadena, CA 91109, USA}

\def\addTAPIR{TAPIR Group, MC 350-17, California Institute of Technology, Pasadena, California 91125, USA}

\author{Michael Holst} 
\address{\addUCSDa} 
\address{\addUCSDb}

\author{Olivier Sarbach}
\address{\addUMICH} 

\author{Manuel Tiglio}
\address{\addUCSDb}
\address{\addUCSDc}

\author{Michele Vallisneri}
\address{\addJPL}
\address{\addTAPIR}



\subjclass[2010]{Primary 83, 35, 65; Secondary 53, 68, 85}

\date{}

\dedicatory{In memory of Sergio Dain}

\begin{abstract}
On September 14, 2015, the newly upgraded Laser Interferometer Gravitational-wave Observatory (LIGO) recorded a loud gravitational-wave (GW) signal, emitted a billion light-years away by a coalescing binary of two stellar-mass black holes. The detection was announced in February 2016, in time for the hundredth anniversary of Einstein's prediction of GWs within the theory of general relativity (GR). The signal represents the first direct detection of GWs, the first observation of a black-hole binary, and the first test of GR in its strong-field, high-velocity, nonlinear regime. In the remainder of its first observing run, LIGO observed two more signals from black-hole binaries, one moderately loud, another at the boundary of statistical significance. The detections mark the end of a decades-long quest, and the beginning of GW astronomy: finally, we are able to probe the unseen, electromagnetically dark Universe by \emph{listening} to it.
In this article, we present a short historical overview of \emph{GW science}: this young discipline combines GR, arguably the crowning achievement of classical physics,
with record-setting, ultra-low-noise laser interferometry,
and with some of the most powerful developments in the theory of differential geometry, partial differential equations, high-performance computation, numerical analysis, signal processing, statistical inference, and data science.
Our emphasis is on the synergy between these disciplines, and how mathematics, broadly understood, has historically played, and continues to play, a crucial role in the development of GW science.
We focus on black holes, which are very pure mathematical solutions of Einstein's gravitational-field equations that are nevertheless realized in Nature, and that provided the first observed signals.
\end{abstract}

\maketitle
\tableofcontents 

\section{Gravitational waves from black holes: a historical milestone} \label{sec:ligo-announcement}

On February 8, 2016, the U.S.\ National Science Foundation announced an upcoming press conference for February 11, modestly titled {\it Scientists to provide update on the search for gravitational waves},  and yet scheduled at the high-profile  
National Press Club in Washington, DC.
By then it was a known secret that the Laser Interferometer Gravitational-wave Observatory (LIGO) had accomplished the first direct detection of gravitational waves, even if the 1,000+ scientist involved in the experiment had amazingly kept details \emph{almost} secret. 

These details were stunning: LIGO had recorded waves from the inspiral and merger of two black holes at a luminosity distance of $\sim 400$ Mpc (a billion light-years). Until then, black-hole binaries were theorized to exist, but never observed, let alone in their final merger phase.
Furthermore, the shape of the gravitational wave implied that both holes were heavier than 25 solar masses, more than was thought possible in the astrophysical theory of binary evolution. 
This would be truly a historic scientific announcement, appropriately marking the 100th anniversary of Einstein's first prediction of the existence of gravitational waves (GWs).

At the press conference, LIGO Laboratory Executive Director David Reitze did not waste time getting to the point. Once he reached the podium he announced simply:
\begin{center}
{\it
Ladies and gentlemen, we have detected gravitational waves. We did it.
}
\end{center}
In this article, we celebrate the LIGO milestone by drawing a brief history of gravitational-wave science, a field that reached its maturity thanks to astonishing progress in experimental physics, but also due to crucial developments in several fields of mathematics and allied disciplines, including differential geometry, partial differential equations (PDEs), high-performance computation, numerical analysis, signal processing, statistical inference, and data science.

\section{Space and time}\label{sec:Intro}

Einstein's theory of general relativity (GR), which is the theory of gravity he introduced to the world in 1915, has many features which are distinct from Newton's gravitational theory.
One such feature is that it is a theory of \emph{space-time}, the geometry of which is not fixed, but instead governed by its own set of field equations.
Rather than being described as a force on a fixed space-time arena, gravity is \emph{embedded} in the geometry of space-time.
In Galileo's and Newton's everyday notion of space and time, we are used to the concepts of a universal time, simultaneity, and a fixed geometry of space.
The latter is governed by our everyday notion of distances, 
$$
\left( \Delta l \right)^2 = \left( \Delta x \right)^2 + \left( \Delta y \right)^2 + \left( \Delta z \right)^2 \, ,   
$$
or in infinitesimal terms,
\be
d l^2 = d x ^2 + d y ^2 + d z ^2 \, .  \label{eq:ds2}
\ee
More accurately, the space of Newtonian physics is a three-dimensional manifold $\cM$ (typically, but not necessarily, $ \mathbb{R} ^3$) equipped with a flat Riemannian metric $g$; that is, a symmetric tensor field of rank $(0,2)$ with vanishing curvature, denoted by $dl^2$ in Eq.~\eqref{eq:ds2}. 
The Riemannian aspect of the Newtonian metric of space implies that at each point it can be brought into diagonal form as in~\eqref{eq:ds2} and that its signature is $(+1,+1,+1)$.
In fact, it can be globally diagonalized, which is equivalent to the assumption, or the consequence, of the existence of a global family of {\it inertial observers}.
It also implies that, without introducing extra structure, any PDEs describing the gravitational field resulting from given sources must be time-independent.
Therefore, even if the sources change in time, these changes are instantaneously transmitted to the gravitational field $\vec{g} = -\vec{\nabla}\phi$. Here, at each instant of time, the gravitational potential is a function $\phi$ on $\cM$ which satisfies {\it Poisson's equation}
\begin{equation}
\nabla^2 \phi (\vec{x}) := g^{ij} \nabla_i \nabla_j \phi (\vec{x}) = 4 \pi G \rho (\vec{x}) \, ,\label{eq:poisson}
\end{equation}
 where the $g^{ij}$ are components of the inverse metric, the symbol $\nabla$ denotes the uniquely associated metric-compatible, torsion free covariant derivative, and $i,j=1,2,3$ denote spatial indices, with repeated indices taken to imply a summation over them (``Einstein summation convention'').
The term $\rho(\vec{x})$ denotes the mass density at the considered point $\vec{x}$ in space, and $G$ is Newton's constant.
As mentioned above, the Riemannian structure of space implies infinite speed of propagation, and explicitly excludes the possibility of gravitational radiation.

In special relativity, one re-examines physical laws in the context of a combined, but fixed, {\it space-time} geometry.
The fixed character of the metric shares its nature with the Newtonian case, but the metric is \emph{Lorentzian}:
\begin{equation}
ds^2 = -c^2dt^2 + dx^2 + dy^2 + dz^2 \, , \label{eq:minkowski}
\end{equation}
with $c$ the speed of light.
Here, the $ds^2$ is simply conventional notation, as the metric is not positive definite: its signature is $(-1,+1,+1,+1)$. 
This approach allows, essentially, without additional structure, to build covariant PDEs which are hyperbolic in nature, and therefore imply a finite speed of propagation (bounded by $c$). 
Intuitively, this opens the room for the existence of radiation, such as in the theory of electromagnetism. 

Despite their fundamentally different approaches to space and time, Newtonian physics and special relativity share an important common feature: their geometry is fixed and, in particular, flat.
That is, there is a fixed arena (space or space-time) where events take place.
The flatness property can be characterized in a number of ways: the Riemann tensor vanishes identically at every point of the manifold, or, equivalently, any vector parallel transported along any closed loop returns to itself.
Examples of intrinsically non-flat two dimensional Riemannian manifolds are spheres, while cylinders are intrinsically flat, despite being embedded in a curved way in $\Real ^3$. 

In order to reconcile gravity with special relativity, in GR these conditions need to be relaxed, and space-time is intrinsically curved and dynamical.
As a consequence, there is, in general, no notion of preferred observers.
The most natural object that describes this type of space-time is a Lorentzian manifold $(\cM,g)$.
That is, a four-dimensional differentiable manifold $\cM$ endowed with a non-degenerate, symmetric rank $(0,2)$ tensor field $g = g_{\mu\nu} dx^\mu dx^\nu$ on $\cM$ whose signature is $(-1,1,1,1)$, as in~\eqref{eq:minkowski}.
The space-time $(\cM,g)$ is required to satisfy the equations that Einstein postulated in 1915, known as {\it Einstein's field equations}, 
\begin{equation}
R_{\mu \nu} - \frac{1}{2} R g_{\mu \nu} = \frac{8 \pi G}{c^4} T_{\mu \nu} \, , \label{eq:einstein}
\end{equation}
where  $R_{\mu \nu}$ is the Ricci curvature tensor, $R$ its scalar ($R:=R_{\mu \nu}g^{\mu \nu}$), $T_{\mu \nu}$ is the stress energy-momentum tensor of any matter fields present.
Greek letters are typically used for space-time indices, $\mu, \nu=0,1,2,3$ and a sum over repeated indices is implicitly assumed.
Equation~\eqref{eq:einstein} has in the vacuum case ($T_{\mu \nu}=0$), as expected, the Minkowski metric~\eqref{eq:minkowski} as a solution.
However, it also has other non-trivial solutions such as black holes, in particular with emission of gravitational energy in the form of waves, as described throughout this article.

What is perhaps less commonly known is that Einstein and Hilbert were in contact throughout the period that Einstein completed the theory of GR, leading to an historical priority dispute.
In fact, on 20 November 1915, five days before Einstein presented his final form of the field equations to the Prussian Academy of Sciences, Hilbert submitted an article in which he formulated the gravitational field equations in terms of a coordinate-invariant variational principle based on what is now called the {\it Hilbert action}~\cite{Hilbert}
\begin{equation}
S[g_{\mu \nu}] = \frac{c^4}{16 \pi G}\int_{\cM} R(g_{\mu \nu})  \sqrt{-|g|} d^4 x,
\label{eq:hilbert}
\end{equation}
where $|g|$ is the determinant of the metric coefficients $g_{\mu\nu}$.
If one computes the functional derivative of $S$ with respect to $g_{\mu \nu}$ (computing its first variation), the Euler-Lagrange system~\eqref{eq:einstein} in the vacuum case is obtained.
The right-hand side of~\eqref{eq:einstein} is obtained by adding to~\eqref{eq:hilbert} a matter term, 
$$
S_{\tt matter} = \int_{\cM} \cL_{\tt matter} \sqrt{-|g|} d^4 x \, , 
$$
the first variation of which leads to $T_{\mu\nu} := -2\delta S_{\tt matter}/\delta g^{\mu\nu}$ in~\eqref{eq:einstein}.
Although a closer analysis revealed that the first proof of Hilbert's paper did not contain the explicit form of Einstein's field equations in terms of the Ricci tensor and its trace~\cite{Corry1997,Sauer1999},
it is Hilbert's work that laid the foundations for the Lagrangian and Hamiltonian formulations of GR.  

A free test particle or ``falling observer''  in GR follows a ``straight'' trajectory in space-time with respect to its geometry; namely, a geodesic. If $\gamma $ is such a trajectory with tangent $t^{\mu}$, then 
\be
t^{\mu}\nabla_{\mu} t^{\nu}  = 0 \, . \label{eq:geodesic}
\ee
That is, the tangent to the {\it worldline} of the observer is parallel transported along itself. In a local coordinate system $\{ x^{\mu} \}$, Eq.~\eqref{eq:geodesic} constitutes a set of four ordinary differential equations, 
\be
\frac{d^2 x^{\nu}}{d \lambda ^2} = 
 -\Gamma^\nu{}_{\mu\sigma} \frac{d x^{\mu}}{d \lambda} \frac{d x^{\sigma}}{d \lambda} \, ,  \label{eq:ode_geodesic}  
\ee
where $\Gamma^\nu{}_{\mu\sigma}$ are the Christoffel symbols generated by the metric and $\lambda$ any (affine) parametrization of the geodesic. From the form of~\eqref{eq:geodesic}, there is one and only one geodesic locally going through any point of space-time with a given direction. If all geodesics have global solutions (for $\lambda \rightarrow \pm\infty$) the space-time is said to be {\it geodesically complete}. However, the physically most interesting space-times in GR, such as the ones describing the gravitational collapse of a star or the expanding universe we live in, can be shown to be geodesically incomplete, according to the famous {\it singularity theorems} by Penrose and Hawking. Despite recent advances, a problem which is still open is to prove that curvature invariants diverge along such incomplete geodesics, yielding a more satisfactory characterization of space-time singularities.

The fact that GR can be formulated as a geometrical theory has a conceptually beautiful consequence: it implies that the laws of physics are invariant under any diffeomorphism of the space-time manifold $\cM$. This in turn implies that the physical laws have precisely the same form in {\it any} local coordinate system ({\it general covariance}), whereas in Newton's theory and special relativity the laws are only invariant among inertial (constant relative velocity) systems. 
The geometry of space-time is built into the covariant derivative associated with the metric, which defines the curvature tensor. The presence of curvature manifests itself in a non-zero acceleration between neighboring geodesics (geodesic deviation). Poisson's equation~\eqref{eq:poisson} and Newton's law for the motion of a test particle in a given gravitational field are recovered from Einstein's field equations~\eqref{eq:einstein} and the geodesic equation~\eqref{eq:geodesic} in the limit of weak gravitational fields and slow motion.

\section{Gravitational waves} \label{sec:GWs}

Early on, GR explained, with no free parameters, several anomalous behaviors within Newton's theory of gravity, such as the precession of the perihelion of Mercury.\footnote{See~\cite{lrr-2006-3} for a review on the status of experimental tests of GR.} From this initial success of the theory, focus shifted to whether the field equations~\eqref{eq:einstein} allowed for non-singular solutions which carried physical energy in the form of gravitational waves (GWs), and what exactly would be meant by that. 

In 1916, Einstein published a paper predicting the existence of GWs~\cite{Einstein:1916cc} by analyzing the weak-field-regime of his field equations, in which the gravitational field is linearized around the flat space-time Minkowski metric $\eta_{\mu\nu}$ given in~\eqref{eq:minkowski}. In modern language, taking a smooth one-parameter family of solutions $g_{\mu\nu}(\lambda)$ of the field equations~\eqref{eq:einstein} with corresponding family of stress energy-momentum tensor $T_{\mu\nu}(\lambda)$ such that $g_{\mu\nu}(0) = \eta_{\mu\nu}$ and $T_{\mu\nu}(0) = 0$ for $\lambda=0$, the linearization of Einstein's field equations can be written as the following constrained wave system (cf. Section~\ref{SubSec:Evolution}):
\begin{equation}
\Box\gamma^{\mu\nu} = \frac{16\pi G}{c^4}\tau^{\mu\nu},\qquad
\nabla_\mu\gamma^{\mu\nu} = 0.
\label{Eq:LinEinstein}
\end{equation}
Here, $\Box = \eta^{\mu\nu}\nabla_\mu\nabla_\nu = -(c^{-1}\partial_t)^2 + \partial_x^2 + \partial_y^2 + \partial_z^2$ is the wave operator with respect to the flat space-time derivative $\nabla$, and
$$
\gamma^{\mu\nu} := \left. \frac{d}{d\lambda} \sqrt{-|g|} g^{\mu\nu}(\lambda) \right|_{\lambda=0},
\qquad
\tau^{\mu\nu} := \left. \frac{d}{d\lambda} T^{\mu\nu}(\lambda) \right|_{\lambda=0}
$$
are the first variations of  $\sqrt{-|g|} g^{\mu\nu}(\lambda)$ and $T^{\mu\nu}(\lambda)$. Therefore, the linearized equations admit wave-like solutions which propagate at the speed of light. In the region where the source is zero, $\tau^{\mu\nu} = 0$, $\gamma^{\mu\nu}$ can be written as superpositions of simple plane wave solutions. Exploiting the coordinate freedom, such plane waves can be expressed as (given here for the case of propagation in the $z$ direction)
\begin{equation}
\gamma_{\mu\nu} dx^\mu dx^\nu = h_+(c t -z) (dx^2 - dy^2) + 2 h_\times(c t-z) dx dy \label{eq:gw-waves}
\end{equation}
with two functions $h_+$ and $h_\times$ parametrizing the two polarizations of the wave.
In these coordinates, known as the ``TT gauge,'' the effect of each GW polarization is to contract \emph{fractionally} the \emph{proper distance} along one axis, while expanding it along the other (these axes being $(x,y)$ for $h_+$, and axes rotated by $45^\circ$ with respect to $(x,y)$ for $h_\times$). In other words, the GWs do not affect the trajectories of freely falling particles,\footnote{The converse is true in the ``laboratory'' gauge, where the GWs act as forces that modify the trajectories, while proper distances between them follow the unperturbed Lorentzian metric.} but change the distances that can be measured between them---for instance, by exchanging pulses of light (see, e.g.,~\cite{Maggiore2000}).

In 1918 Einstein was able to write (albeit with a mistaken factor of $1/2$, later corrected by Eddington) the celebrated \emph{quadrupole formula} for the emission of GWs by a non self-gravitating system in slow motion~\cite{1918SPAW.......154E}:
\begin{equation}
h_{jk} = \frac{2 \, G}{c^4 r}[\ddot{I}_{jk}(t - r/c)]^\mathrm{TT},
\end{equation}
where $\ddot{I}_{jk}$ denotes the second time derivative of the mass quadrupole moment, and $[\cdot]^\mathrm{TT}$ the projection to the transverse--traceless TT frame; $r$ is distance and $t$ is time, so that $t - r/c$ is the retarded time.

For decades it was not clear whether these waves in the metric of space-time had any physical significance.
To begin with, it was thought that they might be a purely coordinate artifact.
At the time, and for several decades subsequent, the concept of covariance and coordinate-invariants had not fully penetrated the minds of relativists (the concept of a black hole was similarly an idea that was difficult to come to terms with, as discussed in Section~\ref{sec:bh}). 
Second, whether GWs could be defined in the full non-linear regime in a theory where space-time itself is dynamic (``waves with respect to what?'') was not at all clear.
Third, it was not known whether it was even possible that there could be non-pathological solutions to the field equations that would admit the existence of such waves. 
Fourth, that any such waves could interact with other forms of energy in a precise, measurable way was unknown.
Finally, whether devices could be designed and built to actually {\em directly detect} these waves seemed like science fiction.   
 
In fact, 
Einstein himself had his own doubts and radically changed his position at times.
These fascinating developments have been researched and documented by Daniel Kennefick~\cite{Kennefick:1142610}. In 1934, Einstein and Nathan Rosen, based on an exact solution to the full (non-linear) field equations, concluded that the theory did not allow for non-singular solutions with GWs carrying energy.  They submitted their manuscript ``{\it Do Gravitational Waves Exist?}''  to Physical Review, with a critical referee report, now known from persistent historical research done by Kennefick to have been from cosmologist Howard Percy Robertson, a report written with exquisite detail and in a very short time.  Einstein replied with a remarkably strong yet not so well known quote:
\begin{displayquote}
{\it
Dear Sir,

We (Mr. Rosen and I) had sent you our manuscript for publication and had not authorized you to show it to specialists before it is printed. I see no reason to address the - in any case erroneous - comments of your anonymous expert. On the basis of this incident I prefer to publish the paper elsewhere.

Respectfully,

Albert Einstein
}
\end{displayquote}

Robertson later met Leopold Infeld, Einstein's new research assistant. Infeld, in turn, exchanged with Einstein, who  had submitted his paper by then to another journal (he never submitted a paper to Physical Review again). Apparently, after his resubmission to another journal, Einstein had also realized or convinced himself of the flaws in the original manuscript, as had Rosen while in Russia (who had reached Einstein by mail about it, but apparently did not manage to do so in time). The main flaw had been seeking for plane GW solutions which turned out to be singular because the proper interpretation was that they actually corresponded to cylindrical waves. 

Einstein radically revised the paper before returning the galley proofs. In particular, he changed its title to ``{\it On Gravitational Waves}''~\cite{Einstein193743}. Robertson wrote to John Torrence Tate, the editor of Physical Review at the time:
\begin{displayquote}
{\it
You neglected to keep me informed on the paper submitted last summer by your most distinguished contributor. But I shall nevertheless let you in on the subsequent history. It was sent (without even the correction of one or two numerical slips pointed out by your referee) to another journal, and when it came back in galley proofs was completely revised because I had been able to convince him in the meantime that it proved the opposite of what he thought. You might be interested in looking up an article in the Journal of the Franklin Institute, January 1937, p. 43, and comparing the conclusions reached with your referee's criticisms.
}
\end{displayquote}

The telling of this story is not meant to undermine Einstein's great physical intuition but, on the contrary, to highlight the level of subtlety in making sense of what gravitational radiation means in a theory of space-time itself, a meaning that it might now be taken for granted as always understood. Also, in some fairness to Einstein, at the time papers in Germany were not peer-reviewed and it was also a rather new practice in Physical Review. Still, Robertson was right. 

The discussion of the concept of gravitational radiation continued for several decades. As an example of the acknowledged level of confusion, the interested reader can go through the proceedings of the second conference organized by the International Committee on General Relativity and Gravitation (to later become the International Society on General Relativity and Gravitation), a seminal and pivotal meeting that took place in Chapel Hill in 1957 (Einstein passed away a few months before the first conference, GR0, which took place in Bern in 1955).  The conference resulted in standard proceedings, but also a report for the Air Force~\cite{ChapelHill}, which had sponsored the meeting (an interesting story by itself). The report has summaries of many of the discussions which took place at the conference. As an example of the discussion on the significance of GWs, John Archibald Wheeler himself is quoted as having said during that meeting
\begin{displayquote}
{\it
How one could think that a cylindrically symmetric system could radiate is a surprise to me (...)
}
\end{displayquote}
\noindent (It is now known that it can.)  The whole topic of GWs was finally put on firm grounds in the sixties after extensive work by Bondi, van der Burg, Metzner, Sachs and Penrose, among others. As it turns out, it is possible to provide an unambiguous (that is, coordinate-invariant) definition of gravitational radiation for {\it asymptotically flat space-times}, those which approach the flat Minkowski metric along outgoing null geodesics in a suitable way. Asymptotic flatness captures the concept of space-times from ``isolated'' or ``bounded" sources. For a review, see~\cite{Frauendiener04}.

\section{Early searches for ripples in space-time}

From the experimental side, in 1969 Joseph Weber announced direct observation of GWs through bar detectors~\cite{Weber:1969bz}. Unfortunately, these results could never be duplicated by other groups in spite of several efforts. Such must have been the perceived potential of such a claimed discovery that a gravimeter was sent to the moon with the hope of detecting modes of the moon excited by GWs \cite{gw-moon}.
Notwithstanding the inability of other groups to reproduce his results, and the NSF cutting his funding in $1987$, Weber continued to work with bar detectors with essentially no funding until his death in $2000$, with the conviction that he had detected not just one but many events. Despite a wide consensus that his measurements did not correspond to GWs, he is recognized as the pioneer of the field of direct GW detection. The sociologist Harry Collins makes interesting observations about scientific interactions in the GW community, and in particular the circumstances around Weber's work \cite{collins2010gravity}.

In 1974, Russell Hulse and Joseph Taylor Jr.\ discovered the binary pulsar PSR B1913+16~\cite{Hulse:1974eb},
whose orbit was later shown by Taylor and Joel Weisberg to shrink in remarkable agreement with the emission of gravitational radiation as predicted by Einstein's quadrupole formula~\cite{1989ApJ...345..434T}.
This discovery, which led to a Nobel prize in 1993 for Hulse and Taylor, was the first clear if indirect demonstration of the existence of GWs.
In fact, the precise timing analysis of PSR B1913+16 (and of other similar pulsars) reflects and demonstrates a broader range of general-relativistic corrections, which are computed in the \emph{post-Newtonian approximation}~\cite{2014LRR....17....2B}: this is an umbrella term for updating Newton's equations perturbatively using a variety of series expansions, most notably with respect to the dimensionless source velocity~$v/c$.

\section{From the curvature of space-time to partial differential equations}

Being the description of a deterministic process, one expects that Einstein's equations~\eqref{eq:einstein} can be cast into two sets: one possibly constraining the initial configurations which are compatible with the theory, and another one describing their evolutions. This is indeed what occurs in  other physical systems, such as Maxwell's equations of electromagnetism, where the electric $\vec{E}$ and magnetic fields $\vec{B}$, in the absence of external charges and currents, are constrained to have vanishing divergence,
$$
\vec{\nabla} \cdot \vec{E} = 0  = \vec{\nabla} \cdot \vec{B},
$$
while their evolution is determined by the laws of Amp\`ere and Faraday
$$
\partial_t\vec{E} = \vec{\nabla}\times\vec{B},\qquad
\partial_t\vec{B} = -\vec{\nabla}\times\vec{E}.
$$
As we discuss next, Einstein's field equations can be split in a similar way. However, although despite their apparent simplicity as written in~\eqref{eq:einstein}, their formulation 
as a Cauchy problem leads to a rather complicated and subtle set of elliptic and hyperbolic equations which are arithmetically expensive to solve for numerically and which have given rise to rich mathematical developments. 
We touch on just a few of these developments which are relevant for this article.

\subsection{The $3+1$ Decomposition of Space-time}

The formulation of Einstein's field equations as a Cauchy problem requires ``breaking'' {\it general covariance} by introducing a {\it foliation} of space-time by three-dimensional hypersurfaces which are typically chosen to be space-like (that is, with a Riemannian intrinsic geometry). Such a foliation is equivalent to the choice of a global time coordinate $t$ on the space-time manifold whose level sets (the ``constant-time slices'') are three-dimensional space-like hypersurfaces $\Sigma_t$. Furthermore, when formulating the problem as a PDE, spatial coordinates $(x^i)$ within each time slice have to be chosen. Clearly, these coordinate choices are highly non-unique, and a ``good'' choice (which is problem-dependent), has been one of the main challenges in the history of mathematical and numerical relativity. This matter is further complicated by the fact that the geometric properties of space-time are not known before actually solving the field equations. 

\begin{figure}
\includegraphics[width=0.8\columnwidth]{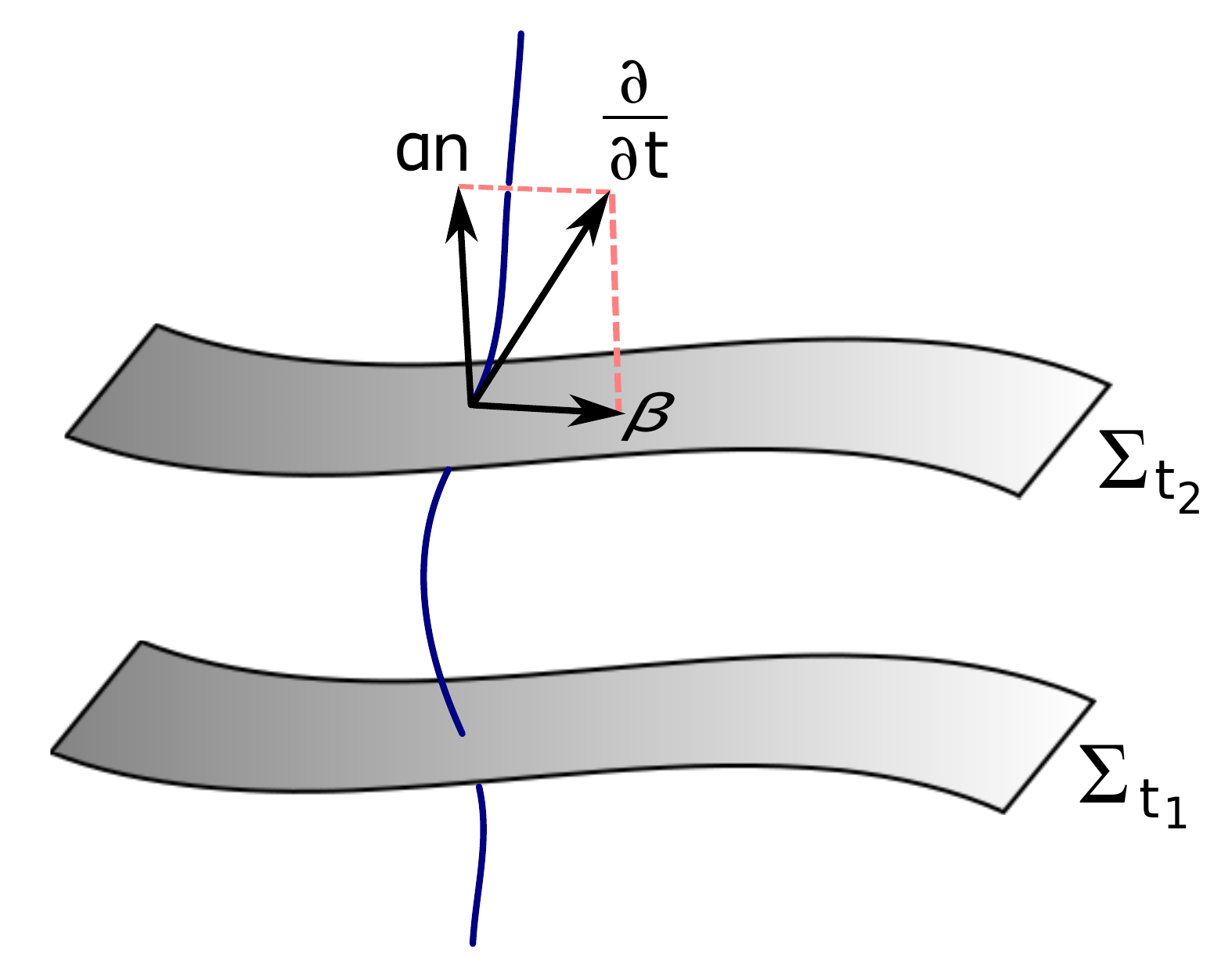}
\caption{Illustration of the $3+1$ decomposition. The blue curve is the worldline of an observer with constant spatial coordinates $(x^i)$. Its tangent vector $\partial/\partial t$ is decomposed into a component parallel to $\Sigma_t$ (the shift vector $\beta$) and one that is orthogonal to it (the lapse $\alpha$ times the normal vector $n$).
Image courtesy of Manuel Morales.}
\label{fig:adm}
\end{figure}

Once a foliation and the spatial coordinates have been chosen, the metric can be decomposed in the following ``$3+1$" form:
\begin{equation}
g = g_{ij} (dx^i + \beta^i dt) (dx^j + \beta^j dt) - \alpha^2 dt^2,
\end{equation}
with $i,j=1,2,3$, and where $g_{ij}$ describes the components of the induced metric on $\Sigma_t$,  where $\beta^i$ are components of the shift vector, and $\alpha$ is the lapse. The lapse is a positive function which only depends on the choice of the time coordinate $t$. Its spatial gradient determines the acceleration $a^\mu$ of the observers through the relation
$$
a^\mu = D^\mu\log\alpha,
$$
where $D$ denotes the induced connection on $\Sigma_t$. The shift determines the  velocity of the observers with constant spatial coordinates $(x^i)$ with respect to the normal observers (those moving perpendicular to the time slices). 
The $3+1$ decomposition of space-time is depicted in Figure~\ref{fig:adm}.

Since the lapse and shift are directly related to the choice of coordinates, they do not carry any information about the dynamical degrees of freedom of the gravitational field, and hence, without any further restrictions on the geometry of the foliation or the evolution of the spatial coordinates, one does not expect the field equations to provide any evolution or constraint equations for them. In other words, these functions 
can be arbitrarily specified. Usually, the most naive choices are not the most appropriate ones. For example, choosing $\alpha = 1$ implies that the normal observers are free falling, which might lead to coordinate-singularities after a finite time of the evolution due to the focusing effect of the gravitational field.

Once the foliation of space-time has been chosen, the evolution and constraint equations are obtained by considering the projections of the field equations~\eqref{eq:einstein} in directions tangent and orthogonal to the space-like hypersurfaces $\Sigma_t$. The evolution equations can be cast as a first-order system for the first and second fundamental forms associated with the time slices, namely the three-metric $g_{ij}$ and extrinsic curvature $k_{ij}$. 
In vacuum, this system reads
\begin{eqnarray}
\hspace*{-0.5cm} \partial_t g_{ij} &\hspace*{-0.1cm}=\hspace*{-0.1cm}& -2\alpha k_{ij} + \pounds_\beta g_{ij},
\label{Eq:gt}\\
\hspace*{-0.5cm} \partial_t k_{ij} &\hspace*{-0.1cm}=\hspace*{-0.1cm}& \alpha(R_{ij}^{(3)} + k k_{ij} - 2k_i{}^l k_{lj}) - D_i D_j\alpha
 + \pounds_\beta k_{ij},
\label{Eq:kt}
\end{eqnarray}
with $i,j=1,2,3$, and
with $R_{ij}^{(3)}$ the Ricci tensor associated with $g_{ij}$, $k = g^{ij} k_{ij}$ the trace of the extrinsic curvature and $\pounds_\beta$ denoting the Lie-derivative operator with respect to the shift vector. 
With $g$ and $k$ symmetric tensors, this represents 12 equations for the 12 components of $g$ and $k$, with the equations being first-order in time and second-order in space.
The four constraint equations on the 12 degrees of freedom are
\begin{eqnarray}
g^{ij} R_{ij}^{(3)} + k^2 - k^{ij} k_{ij} &=& 0,
\label{Eq:HamConstr}\\
D^j k_{ij} - D_i k &=& 0,
\label{Eq:MomConstr}
\end{eqnarray}
and are direct consequences of the Gauss-Codazzi-Mainardi conditions which are required for a $3$-manifold to arise as sub-manifold of a $4$-manifold.
If matter and/or energy sources are present, then the 12 evolution equations~\eqref{Eq:gt}--\eqref{Eq:kt} and the four constraint equations~\eqref{Eq:HamConstr}--\eqref{Eq:MomConstr} contain additional terms for this non-vaccum setting.

As anticipated, there are no evolution equations for the lapse and shift, nor 
are they restricted by the constraints. 
Furthermore, it should also be noted that the evolution equations~\eqref{Eq:gt}--\eqref{Eq:kt} are not unique, since they can be modified using the constraints. Such freedom has proven crucial in mathematical and numerical relativity, such as in the formulation of a well-posed evolution problem, or being able to simulate binary black holes on a computer.

The evolution and constraint equations can also be derived from a Lagrangian through a variational principle; one incorporates the lapse $\alpha$ and shift $\beta^i$ into the Hilbert action~\eqref{eq:hilbert} for the space-time metric $g_{\mu \nu}$, and then exploits the foliation by splitting the space-time action integral into a time action integral of a spatial Lagrangian. The result is a second-order constrained evolution system for the unknown spatial metric $g_{ij}$. In 1959, Arnowwitt, Deser, and Misner~\cite{ADM59} 
developed an analogous Hamiltonian formulation that produces a first-order constrained hyperbolic system for $g_{ij}$ and a conjugate momenta $\pi^{ij}$ which is directly related to the extrinsic curvature $k_{ij}$. In their formulation, the lapse and shift appear as Lagrange multipliers in the Hamiltonian, and the constraint equations are produced by varying the Hamiltonian with respect to the lapse and shift, giving rise to the so-called {\em Hamiltonian} and {\em momentum} naming of the two constraint equations~\eqref{Eq:HamConstr} and \eqref{Eq:MomConstr}.

\subsection{The Evolution Problem}
\label{SubSec:Evolution}

The study of the initial-value problem for Einstein's equations was pioneered by the French school. Darmois already in 1927~\cite{darmois1927memorial} showed existence and uniqueness to the Einstein evolution equations for the case of analytic initial data. This is too restrictive and at odds with causality, though, since in the analytic case the initial data would be entirely determined by their value in any (arbitrary small) open set.

In 1939 Liechnerowicz, a student of Darmois (who had also worked with Cartan) presented in his Ph.D.\ thesis a geometric version of GR and of the  initial-value problem~\cite{lichnerowicz1955theories}. This was followed up, in turn, by Choquet-Bruhat in 1952 with a proof of local existence and uniqueness of solutions of the initial value problem only assuming that the initial data is smooth enough (i.e. relaxing the analyticity condition)~\cite{cB52}. This analysis used harmonic coordinates for the space-time, defined as satisfying
\begin{equation}
\nabla^{\mu}\nabla_{\mu} x^{\nu} = 0 \, ,  \quad \nu = 0,1,2,3,
\label{eq:harmonic}
\end{equation}
for which the Einstein equations turn into a set of ten quasilinear wave equations. In the weak field regime the linearized system reduces precisely to the constrained wave system~\eqref{Eq:LinEinstein}.

More recently, harmonic coordinates~\eqref{eq:harmonic} have even been used to prove the global non-linear stability of the Minkowski space-time by Lindblad and Rodnianski~\cite{hLiR05}. Although a somehow stronger result had already been obtained much earlier in the pioneering work by Christodoulou and Klainerman~\cite{Christodoulou93}, the fact that harmonic coordinates could be used to establish global results came as a surprise, since it had been conjectured early on that in general~\eqref{eq:harmonic} would not yield globally defined coordinates even for space-times near the Minkowski solution.

In the strong field regime, one does expect harmonic coordinates to eventually break down. One possible solution for evading coordinate pathologies is to add low order source terms to Eq.~\eqref{eq:harmonic}, which do not affect the well-posedness of the initial-value problem. A variation of this formulation, with source terms in the spirit proposed by Helmut Friedrich years in advance~\cite{Friedrich85}, was one of the ingredients used about 30 years later by Frans Pretorius to perform the first long-term stable simulation of binary black holes~\cite{Pretorius:2005gq}.  

There are, of course, many other ways of posing the initial-value problem in GR which are based on different coordinate choices, different choices of variables and different ways of using the constraints in order to modify the evolution equations. 

\subsection{The Constraint Equations}

The four constraint equations~\eqref{Eq:HamConstr}--\eqref{Eq:MomConstr},
with additional terms included when there are matter and energy sources,
form an underdetermined system of equations for the 12 degrees of freedom (the components of the symmetric two index tensors $g$ and $k$).
Similar to the situation in electromagnetism (Maxwell's equations), the constraints within the Einstein equations restrict the class of initial data allowed for the evolution problem in GR.
If the constraint equations are written schematically as $C_\alpha = 0$, $\alpha = 0,1,2,3$, then if the evolution equations are satisfied, the constraint variables $C_\alpha$ themselves satisfy a homogeneous evolution equation.
In the Maxwell case this evolution system is trivial, $\partial_t(\vec{\nabla} \cdot \vec{E}) = \partial_t(\vec{\nabla} \cdot \vec{B}) = 0$, but in GR it involves non-trivial speeds of propagation.
In the linearized case, Eq.~\eqref{Eq:LinEinstein} together with the fact that $\nabla_\mu\tau^{\mu\nu} = 0$ imply that
$$
\Box C^\nu = 0,\qquad C^\nu := \nabla_\mu\gamma^{\mu\nu},
$$
so that the constraint variables satisfy a wave system on their own.
Therefore, if the initial data satisfy the constraints, one can show that they are preserved during evolution.
While this is true for exact solutions of the evolution equations, it fails to hold for numerical (and other) approximations.
In numerical GR, one thus distinguishes between {\it free evolution} (explicitly solving only the evolution equations, and monitoring the growth of constraint violation as the system evolves~\cite{NR:Book}), and {\it constrained evolution} (explicitly solving both the evolution and constraint equations, using techniques such as constraint projection~\cite{Holst:2004wt}).

Working on initial-value problems in GR (either proving existence theorems or explicitly producing numerical solutions) involves solving the constraint equations.
The constraints have been studied by mathematicians as a stand-alone PDE system since the 1940's~\cite{L44}; this has been one of the research areas in GR that mathematicians have made substantial contributions.
For example, it has been shown that the set of solutions of the constraints forms an infinite-dimensional manifold, except at a small set of points~\cite{FMM80}; this implies that for a generic point from the solution set, it should be possible to specify a chart of this manifold covering a neighborhood of that point, effectively ``parameterizing'' the possible initial data.
The most useful tool for building such parameterizations, and for developing a more complete understanding of the constraint equations, has been the so-called \emph{conformal method}.
The conformal method provides a practical computational procedure for solving the constraint equations, and it also forms the basis for other theoretical tools such as \emph{gluing techniques} (see~\cite{C00,CS03,CD03,CIP05,Mazz09} and the references therein).
Applications of the conformal method include construction of initial data for black holes~\cite{DJK06}, binary systems of black holes and stars~\cite{Cook:2000vr}, gravitational radiation~\cite{BoY80}, and many other models.

\subsection{The Conformal Method}
\label{sec:conformal}

The conformal method was proposed by Lichnerowicz in 1944~\cite{L44}, and then substantially generalized in the 1970s by York~\cite{Y73}, among other authors.
The method is based on a splitting of the initial data $\hat{g}_{ij}$ (a Riemannian metric on a space-like hypersurface $\Sigma_t$) and $\hat{k}_{ij}$ (the extrinsic curvature of the hypersurface $\Sigma_t$) into eight freely specifiable pieces, with four remaining pieces to be determined by solving the four constraint equations.
The pieces of the initial data that are specified as part of the method are a spatial background metric $g_{ij}$ on $\Sigma_t$ (six free functions), and a transverse, traceless tensor $\sigma_{ij}$ (two free functions).
The two remaining pieces of the initial data to be determined by the constraints are a scalar conformal factor $\phi$ and a vector potential $w^i$.
The full spatial metric $\hat{g}_{ij}$ and the extrinsic curvature $\hat{k}_{ij}$ are then recovered from $\phi$, $w^i$, and the eight specified functions from the expressions:
$
\hat{g}_{ij} = \phi^4 g_{ij},
$
and
$
\hat{k}^{ij} = \phi^{-10}[\sigma^{ij} + (\mathcal{L}w)^{ij}]
    + \frac{1}{3} \phi^{-4} \tau g^{ij}.
$
This transformation has been engineered so that the constraints~\eqref{Eq:HamConstr}--\eqref{Eq:MomConstr} reduce to coupled PDEs for $\phi$ and $w^i$ with standard elliptic operators as their principle parts:
\begin{align}
- 8 \Delta \phi 
    + R \phi
    + \frac{2}{3} \tau^2 \phi^5
    - (\sigma_{ij}+(\mathcal{L}w)_{ij})^2
      \phi^{-7}
&=0,
\label{eqn:ham_conf}
\\
- \nabla_i (\mathcal{L}w)^{ij} + \frac{2}{3} \phi^6 \nabla^j \tau 
&=0.
\label{eqn:mom_conf}
\end{align}
Here, $\Delta$ is the Laplace-Beltrami operator with respect to the background metric $g_{ij}$, $\mathcal{L}$ denotes the \emph{conformal Killing operator}
$(\mathcal{L}w)_{ij} = \nabla_i w_j + \nabla_j w_i - \frac{2}{3} (\nabla_k w^k) g_{ij}$, and $\tau= \hat{k}_{ij}\hat{g}^{ij}$ is the trace of the entrinsic curvature.
A detailed overview of the conformal method, and its variations, may be found in the 2004 survey in Ref.~\cite{BI04}.

Note that if the hypersurface $\Sigma_t$ has constant mean extrinsic curvature (known as the CMC case), then the term in~\eqref{eqn:mom_conf} involving $\nabla^j \tau$ vanishes, and the two equations decouple; one can first solve~\eqref{eqn:mom_conf} for $w$, and then plug $w$ into~\eqref{eqn:ham_conf} and solve for $\phi$.
The conformal method was initially used in this decoupled form in numerical relativity (cf.~\cite{Cook91,CoTe99}), but methods for the coupled system~\eqref{eqn:ham_conf}--\eqref{eqn:mom_conf} were also developed~\cite{Cook:2000vr,BeHo96,Hols2001a,Pfei04}.
However, mathematical proofs of existence and uniqueness of solutions were limited to the decoupled case through 1995~\cite{MuYo73,JI95}.
Then in 1996, it was shown~\cite{IM96} that if $\Sigma_t$ has nearly-constant mean extrinsic curvature (the \emph{near-CMC case}), then some CMC results (for compact manifolds) could be extended to the near-CMC case where the equations are coupled.
Between 1996 and 2007 a number of such extensions appeared, including results for Euclidean~\cite{CBIY00} and asymptotically hyperbolic manifolds~\cite{AC96}.
The CMC case was also further developed, including results for open manifolds with interior ``black hole'' boundary models~\cite{sD04,dM05b}, results allowing for ``rough'' data~\cite{DMa06,yCB04}, and other results.

In 2008--2009, it was shown~\cite{HNT07a,HNT07b,dM09} that the near-CMC assumption could be avoided; if other parts of the data ($\sigma$ and matter sources present) are not too large, then there exists a solution to~\eqref{eqn:ham_conf}--\eqref{eqn:mom_conf} for arbitrarily prescribed mean extrinsic curvature.
These ``far-from-CMC'' results involved an influx of ideas from PDE and geometric analysis, including techniques used in mathematical elasticity~\cite{sD06}.
While initially for compact manifolds, these results have been extended to asymptotically Euclidean manifolds~\cite{DIMM14,HoMe14a,BeHo14a}, manifolds with asymptotically cylindrical or periodic ends~\cite{CM12,CMP12}, compact manifolds with interior black hole boundaries~\cite{HoTs10a,HMT13a,Dilts:2013}, rough data and metrics~\cite{HNT07b,BeHo14a}, and other settings.
More complete overviews of known results through 2011 include~\cite{CGP10a,CoPo10,Choq09}.

It was initially hoped that the new results that began to appear in 2008 would lead to a solution theory for the non-CMC case that would mirror the CMC case; however, the story has become much more interesting.
The new existence results did not come with uniqueness, and there had already been growing numerical evidence that multiple solutions were possible in the non-CMC case~\cite{PY05,BOP07,HoKu09a}.
A careful analysis in 2011~\cite{M11} confirmed this feature of the conformal method in the non-CMC case.
This is quite undesirable for many reasons, and it has led to a new influx of tools, such as analytic bifurcation theory and closely related numerical continuation methods, to try understand what is going on.
Later we will describe some of the interesting mathematical problems this new activity has generated. 


\subsection{The Initial-Boundary Value Formulation}
\label{sec:ibvp}

\begin{figure}
\includegraphics[width=0.7\columnwidth]{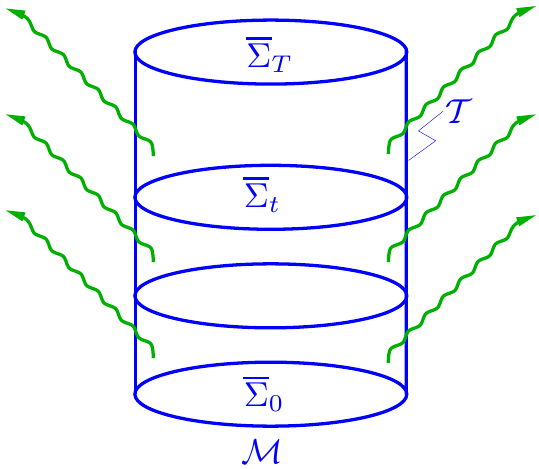} \\
\caption{Illustration of the space-time region $\cM := [0,T]\times \overline{\Sigma}$ on which the IBVP is posed. In numerical relativity one tries to formulate appropriate ``radiative" boundary conditions at the artificial outer boundary ${\cT}$.}
\label{fig:ibvp}
\end{figure}

For numerical applications, one usually does not consider the evolution problem on the whole spatial domain, which is unbounded for typical applications (including the modeling of binary black holes); one rather works on a truncated domain $\overline{\Sigma}$ with artificial (inner or outer) boundaries where appropriate boundary conditions should be specified (see Figure~\ref{fig:ibvp}).
This leads to the consideration of an initial-boundary value problem (IBVP) for Einstein's field equations~\eqref{eq:einstein}. There are several issues that render this IBVP much more difficult than in other physical problems, which are in particular due to the presence of constraints with non-trivial speeds of propagation,  the diffeomorphism invariance of the theory, and the difficulty of obtaining a local characterization of in- and outgoing gravitational radiation in the full non-linear theory. The latter is directly related to the difficulties relativists had had in defining GWs in an unambiguous way, see the discussion in Section~\ref{sec:GWs}.

The first well-posed IBVP for the full non-linear field equations in vacuum was formulated by Friedrich and Nagy in 1999~\cite{Friedrich99}, based on a tetrad description of the gravitational field and the use of the theory of symmetric hyperbolic systems with maximal dissipative boundary conditions. A generalization of this work to the quasi-linear wave system for the metric fields $g_{\mu\nu}$ obtained using harmonic coordinates was given a few years later~\cite{Kreiss:2006mi,Kreiss:2008ig}, 
based on Heinz-Otto Kreiss'  {\it strong well-posedness} concept, which consist of an $L^2$-type estimate which controls both the bulk- and the boundary-norm of the solution. Although these results allow one to construct a unique solution of Einstein's field equations with given initial and boundary data on a space-time region of the form $[0,T]\times\overline{\Sigma}$ for $T > 0$ sufficiently small, a geometric characterization of the boundary data on the boundary surface ${\cT} := [0,T]\times \partial\overline{\Sigma}$ is still missing, and much less is known about global existence of solutions for these problems. A further issue is the specification of ``radiative" type boundary condition at ${\cT}$.
So far, this has only been addressed in some detail at the linearized level. For a recent review on these topics, see~\cite{Sarbach:2012pr}.

\newpage

\section{Black holes}   \label{sec:bh}
\subsection{Stationary Black Holes and the No Hair Theorem}
\label{sec:no-hair}

In late 1915, just a few months after Einstein presented the final form of the gravitational field equations, Karl Schwarzschild found the first non-trivial (i.e. other than the Minkowski space-time) exact solution to the vacuum field equations. Einstein himself was positively surprised that an exact solution could be found at all.  The circumstances under which this happened are remarkable, if not heroic. Schwarzschild derived his solution while serving for the German Army in World War I, which he had joined in 1914, stationed on the Russian front and suffering from a rare and painful skin disease for which there was no cure at the time. During that period, he managed to write three papers. He died in 1916, two months after having been freed from military duty due to illness. 

The Schwarzschild solution is given by
\begin{equation}
g = -\left( 1 - \frac{2m}{r} \right) c^2 dt^2 + \frac{dr^2}{1 - \frac{2m}{r}}
 + r^2\left( d\vartheta^2 + \sin^2\vartheta\, d\varphi^2 \right),
\label{eq:Schwarzschild}
\end{equation}
with $m$ a real constant. It is spherically symmetric and static. In 1923, the mathematician George David Birkhoff proved that any spherically symmetric solution of the vacuum equations must be static, which implies that the Schwarzschild metric is the most general spherically symmetric vacuum one. Note that for $r\to \infty$ the metric~\eqref{eq:Schwarzschild} converges to the flat space-time metric~\eqref{eq:minkowski} (written in spherical coordinates), and in this sense $g$ is asymptotically flat. For finite $r \geq R > 2m$, the metric~\eqref{eq:Schwarzschild} describes the exterior of a spherically symmetric matter distribution of mass $M = mc^2/G$ which is confined to a sphere of radius $R$. Due to Birkhoff's theorem, a spherically symmetric pulsating star cannot emit gravitational radiation. 

\begin{figure}
\includegraphics[width=0.75\columnwidth]{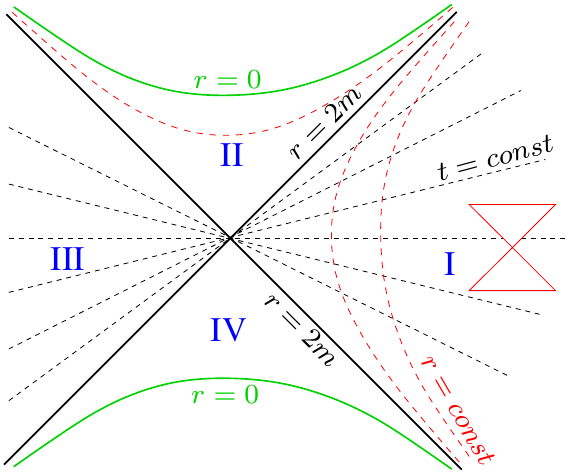}
\caption{Kruskal diagram showing the causal structure of the maximally extended Schwarzschild space-time. In this diagram, the angles $(\vartheta,\varphi)$ are suppressed, and the radial null geodesics are straight lines with slopes $1$ or $-1$, so that the light cones look the same as in Minkowski space-time at each point. Region I corresponds to the points with $r > 2m$ in the original coordinates $(t,r)$, and region II is the black hole region. Regions III and IV are isometric to regions I and II, and describe, respectively, another asymptotic flat end (which is causally disconnected from region I) and a white hole.}
\label{fig:Kruskal}
\end{figure}

The Schwarzschild metric as written in~\eqref{eq:Schwarzschild} seems to have a singularity at $r=2m$. For a long while it was thought that it corresponded to a physical singularity and was referred to as the {\it Schwarzschild singularity}. However, it turns out to be a pure coordinate effect, with all curvature invariants remaining finite. A coordinate change reveals that the metric is perfectly regular at the Schwarzschild radius and that the set $r = 2m$ is a null surface describing the {\it event horizon} of a {\it black hole}. Objects or signals emanating within the region $r<2m$ cannot escape to the outside region $r > 2m$ without exceeding the speed of light. This is why black holes were given their name: their interior cannot be observed, in particular by electromagnetic means. The global structure and geometric understanding of the Schwarzschild space-time as a black hole came about in the late fifties from work by physicist David Finkelstein and mathematician Martin David Kruskal (see Figure~\ref{fig:Kruskal}).

Therefore, Schwarzschild's metric was not only the first exact solution to the field equations, but also one of major historical significance, since it predicted the existence of black holes in the Universe! 
The fact that it took almost 50 years to generalize the Schwarzschild solution to the physically more realistic rotating case is another manifestation of the complexity of Einstein's field equations. A rotating generalization, depending on the mass parameter $m$ and a new  parameter $a$ related to the spin of the black hole, was eventually found by mathematician Roy Kerr in 1963 using a particular ansatz for the metric.  
Later, an alternative derivation based on a systematic study of stationary, axisymmetric solutions of the vacuum Einstein equations was provided. In cylindrical-like coordinates $(t,\varphi,\rho,z)$ the field equations can be reduced to a non-linear elliptic equation, the so-called Ernst equation,
\begin{equation}
\frac{1}{\rho}\partial_\rho\left(\rho\partial_\rho E \right) + \partial_z^2 E 
 = \frac{(\partial_\rho E)^2  + (\partial_z E)^2}{\mbox{Re}(E)},
\label{eq:Ernst}
\end{equation}
for a complex-valued function $E$ of the two variables $(\rho,z)$. Once the Ernst equation has been solved, the metric coefficients $g_{\mu\nu}$ can be obtained from $E$ by quadrature.

In addition to providing a systematic derivation for the Kerr metric, the Ernst equation has played a prominent role in the uniqueness theorems for rotating black holes. Under suitable regularity conditions, one can prove that the exterior of any stationary, asymptotically flat vacuum black hole space-time $(M,g)$ with connected, non-degenerate event horizon is isometric to the exterior of the Kerr space-time. A precise formulation of this uniqueness theorem, along with generalizations to non-vacuum space-times and open problems, can be found in~\cite{lrr-2012-7}.
The most important open issue is the question of whether or not the analyticity assumption on the space-time, which is part of the regularity assumptions in the current formulation, can be relaxed.

The physical implication of the uniqueness theorem is that equilibrium vacuum black holes are very ``simple'' objects, uniquely characterized by their mass and angular momentum (more precisely, by the Kerr space-time). This leads to the popular statement that black holes ``have no hair''. This should be contrasted with the case of rotating neutron stars, where one needs to deal with microphysics or equations of state describing the matter (much of which is largely unknown) which can influence the external structure of space-time.

When two black holes collide, it can be shown that they merge and form a new black hole which cannot be further split into two or more black holes. One might wonder what the evolution of the final black hole looks like. If it decays to a stationary state, according to the no hair theorem it has to be a member of the Kerr family.  Whether that decay takes place or not is related to the issue of black hole stability, described below.

\subsection{The Black Hole Stability Problem}
\label{sec:bh-stability}

Another example of the complexity of Einstein's field equations is that, even after $50$ years since the discovery of the Kerr solution, its stability is still an open mathematical question. The Kerr stability problem is the following: given initial data $(g_{ij},k_{ij})$ which consists of a small perturbation of the data corresponding to a Kerr black hole with initial parameters $(m_i,a_i)$, does this data give rise to a global (in time) solution which settles down to a Kerr black hole with final parameters $(m_f,a_f)$ close to the initial parameters?

This is not just an academic problem, since (as described above) it is related to the expected final state of the collision of two black holes. 
When searching for gravitational waves through matched filtering as described in Section~\ref{sec:data-analysis}, the answer to this question impacts the types of scenarios that should be considered when building catalogs of GW templates. 
Despite rapid recent progress towards the stability problem, so far only results in the linearized case are available. 

In 1957, the first perturbative analysis of the Schwarzschild metric was published by Regge and Wheeler, using mode analysis~\cite{Regge:1957rw}. They found that under certain, physically inspired, boundary conditions at the ``Schwarzschild singularity'' and at infinity, there were no growing modes. In 1989, Bernard Whiting proved mode-stability of the Kerr solution~\cite{Whiting:1989vc}. Although mode stability is a necessary condition for stability, it does not imply on its own that solutions stay bounded for all time.

Moving beyond mode stability, a toy model for black hole stability consists of studying the behavior of the solutions to the wave equation on the fixed background geometry of a Schwarzschild or Kerr black hole. Although the coupling with gravity is neglected, if there was an instability under these conditions, it would be a {\em very} strong indication that when such an external field is coupled to gravity, the black hole would be unstable. 

In the Schwarzschild case, in 1987 Bernard Kay and Robert Wald showed that for smooth, compactly supported initial data on a Cauchy surface, the solution to the scalar wave equation on a Schwarzschild background is uniformly bounded in the exterior region at all times~\cite{Kay:1987ax}. Here, the key innovation was to allow for the first time for initial data that is not assumed to vanish at the event horizon. More recently, work by Mihalis Dafermos, Igor Rodnianski and collaborators considerably strengthened the results by Kay and Wald by proving decay of solutions to the scalar wave equation for the more general case of a Kerr black hole background, thus providing the first proof of asymptotic stability~\cite{Dafermos:2014cua}.

Regarding gravitational perturbations, Dafermos, Holzegel and Rodnianski, just a few months prior to this writing, have presented a proof showing that solutions of the linearized Einstein vacuum equations around a Schwarzschild black hole arising from regular initial data decay to a linearized Kerr metric~\cite{Dafermos:2016uzj}, 
with the decay being inverse-polynomial with respect to the time function of a suitable foliation.

\section{Origins of numerical analysis and numerical relativity}

In parallel to all these theoretical and mathematical developments, in the 1940's, Crank and Nicholson, von Neumann, and collaborators, presented the first stability analyses for numerical solution of time dependent PDE problems. The early 1950's were marked by the first pure and applied mathematical computations on the first general purpose electronic computer, the ENIAC~\cite{metropolis1980history}, including the first weather modeling calculation on an electronic computer, by John von Neumann and colleagues~\cite{TELLUSB12384}. 

Bryce DeWitt is perhaps best known for his work on quantum gravity. While at Lawrence Livermore National Lab, however, where he worked between 1952 and 1955, he became involved in numerical hydrodynamical calculations. At the 1957 GR1 conference, DeWitt and Charles Misner suggested the use of computers to numerically solve Einstein's equations.  Misner summarized one of the sessions in the following way:
\begin{displayquote}
{\it
First we assume that you have a computing machine better than anything we have now, and many programmers and a lot of money, and you want to look at a nice pretty solution of the Einstein equations. The computer wants to know from you what are the values of $g_{\mu \nu}$ and  $\partial g_{\mu \nu} /\partial t$ at some initial surface, say at ${t=0}$.  Now, if you don't watch out when you specify these initial conditions, then either the programmer will shoot himself or the machine will blow up. In order to avoid this calamity you must make sure that the initial conditions which you prescribe are in accord with certain differential equations in their dependence on $x,y,x$ at the initial time. These are what are called the ``constraints'' $(\ldots)$.
}
\end{displayquote}
Through the years these concerns proved to be even more serious in practice in numerical simulations of the Einstein evolution equations. Even if solving for the initial constraints exactly or within machine precision, the behavior of the evolved solutions under constraint violations (which are always present due to finite machine precision or truncation errors) greatly depends on how the equations are written~\cite{Kidder:2001tz}. This was for decades one of the most important problems in numerical relativity, with devised mechanisms to overcome it such as making the space of constraint-satisfying solutions an attractor~\cite{Brodbeck:1998az} or using constraint projection-based methods~\cite{Holst:2004wt}.  

In 1957, Susan Hahn completed her Ph.D.\ thesis, {\it Stability criteria for Difference Schemes}~\cite{CPA:CPA3160110207}, under the supervision of mathematician Peter Lax at NYU. That year coincided with von Neumann's death, who had pioneered the field of numerical methods for PDEs and their applications; however, he did not witness the birth of numerical relativity. 

By 1958, Hahn was already working at IBM, which eventually turned out to be very useful (if not crucial) for being able to carry out the first simulations of colliding black holes. The reason for this was two fold: her expertise on numerically solving PDEs and her access to supercomputers at IBM. Already by the 28th of January in 1958, at the Stevens Meeting in New Jersey, Richard Lindquist, by then a student of Wheeler, reported on his ongoing work on numerically computing the gravitational radiation emitted from an encounter of two black holes~\cite{BrillStevens}. Part of the interest was to measure the acceleration between black holes, but also trying to shed light, {\em by numerical means}, on the by then ongoing {\em theoretical} discussion and research about the concept itself of GWs. In more modern terms, one could rephrase that effort as ``exploratory analysis''.

The research trajectories of applied mathematician Hahn and general relativist Lindquist were about to collide.

\section{The first simulations of head-on black hole collisions}

The first numerical relativity simulation of two black holes was carried out by an applied mathematician and a physicist---a synergy between basic and applied mathematics, GR and what is now referred to as scientific computing or computational science, that would persist over time. 

In 1964, Hahn and Lindquist published the first (super) computer simulation of binary black holes~\cite{1964AnPhy..29..304H}, evolving Misner's initial data for the axisymmetric head-on collision of  two non-spinning black holes initially at rest. 
The evolution equations were written as a first order (in time and space) system of 12 2+1 (2 space dimensions plus time) equations. At the discrete level, second order centered spatial differences and an explicit forward (Friedrich's) scheme for time integration were used. This effort constituted not only a historical landmark in the numerical simulation of Einstein's equation, but Ref.~\cite{1964AnPhy..29..304H} also contains a detailed discussion of numerical stability, to an extent that is comparable to work in numerical relativity in the present day.  
In terms of accuracy, the 2-dimensional spatial mesh consisted of modest $51 \times 51$ points. The computations were done on an IBM 7090 supercomputer, with a reported speed of 4 minutes per time step, for a total evolution time of 50 steps. These simulations would become unstable very quickly, but they constituted the foundations of what decades later would become one of the cornerstones of GW science. 

The IBM 7090 is reported to have had the following speed~\cite{IBM7090}: {\it The 7090 can perform any of the following operations in one second: 229,000 additions or subtractions, 39,500 multiplications, or 32,700 divisions}. The machine cost around USD $\$3,000,000$ and could be rented for about USD $\$60,000$ a month. In comparison, as of this writing a relatively high-end smartphone is capable of {\em orders of magnitude} more FLOPs (floating point operations per second). For example, tests on standard linear algebra benchmarks (LINPACK) currently give a few hundred of MegaFLOPs (millions of FLOPs) on one thread to more than a GigaFLOPs (billions of FLOPs) on some multithread, quad-core smartphones. 
 
In the early 70's, DeWitt took aim at again numerically tackling the equivalent of the two body problem in GR. In a milestone piece of work, a student  of DeWitt, Larry Smarr, revisited the axisymmetric head-on collision of two black holes. His simulations were restricted to  relatively small initial separations between the black holes. However, they were long enough and had enough accuracy to ``extract'', for the very first time, GWs from computer-generated space-times~\cite{NYAS:NYAS569}. The aspects which led to such improvements were not just raw computational power, but a much better understanding of the problem at the continuum. In, particular concerning coordinate conditions that would avoid coordinate singularities or difficulties to simulate unphysical small or multi-scale structures. In conjunction with work by Kenneth Eppley~\cite{eppley75}, this effort by Smarr is recognized as the second milestone in the numerical simulation of black holes. 

From a chronological perspective, as discussed in Section~\ref{sec:ibvp}, in the early seventies Kreiss, Gustafsson and Sundstrom introduced a well-posedness/numerical stability analysis framework for time-dependent initial-boundary-value problems~\cite{GKS}. However, it would not reach the field of numerical relativity until three decades later. 

\section{From numerical relativity to supercomputing }
Remarkably enough, as documented in the Lax report~\cite{LaxReport}, by $1982$ supercomputers in the United States were only available to industry and federal labs (in the latter case mostly for classified projects). Many of the supercomputer simulations by US researchers in academia were carried out in Germany, for example. Smarr had used for his black hole simulations supercomputer resources from Livermore Lab through James Ricker Wilson, who worked on weapon design but also pioneered the field of numerical relativistic hydrodynamics.

In 1983, 
Smarr submitted to NSF the first \emph{unsolicited} proposal to be funded by the agency,  {\it A Center for Scientific and Engineering Supercomputing}. Also known as the ``Black Proposal'' for the color of its cover~\cite{BlackProposal}, 
it resulted in the first network of supercomputer centers in the US available to academia: the Cornell Theory Center, the National Center for Supercomputer Applications (NCSA) at the University of Illinois in Urbana-Champaign, the Pittsburgh and San Diego Supercomputer Centers, and the John von Neumann Center at Princeton. The current ``incarnation'' of this first network is XSEDE (Extreme Science and Engineering Discovery Environment), a consortium of $16$ supercomputers across the US. 

We have highlighted the role and importance of mathematical theory, numerical analysis and computational science on numerical relativity from the birth of the field and throughout the years. The Black Proposal and its award is an example of the opposite: a numerical relativist taking the lead in making supercomputing accessible to {\em any} academic institution and research area in the US.

\section{Interferometric detectors}

In 1972, Rainer Weiss distilled early ideas to measure GWs with km-scale laser interferometers in a detailed (if hard to find) technical report~\cite{Weiss}. The report identified and analyzed the crucial sources of noise, their impact, and mitigation. Along with the contributions of LIGO co-founders Ronald Drever and Kip Thorne, Weiss' work provided the foundation for the design of LIGO, the U.S.\ GW observatory that would be realized in almost identical facilities in Hanford (Washington) and Livingston (Louisiana).
The French--Italian Virgo collaboration implemented a similar design in Italy.

These first-generation detectors took data (but reported no detections) between 2000 and 2010, then both were upgraded to more daring and sensitive \emph{advanced} configurations. Advanced LIGO performed its first (and ultimately successful) observing run in late 2015, while Advanced Virgo is, as we write, in the final stages of commissioning.
Smaller, less sensitive interferometers were built in Germany (GEO600) and Japan (TAMA), and were operated through the early 2000s. Japan is now building its own km-scale interferometer (KAGRA); India is planning to build a 4 km LIGO-like infrastructure to house an already-built third Advanced LIGO detector (LIGO-India). The primary motivations to build an international network of detectors are to improve the localization of GW sources, which is performed essentially by triangulation; to extract information about the polarization of the wave; and to provide greater detector ``uptime''.

LIGO-like detectors measure GWs by using laser interferometry to monitor the differential changes in length along two perpendicular arms.
As explained in Section~\ref{sec:GWs}, GWs produce a \emph{fractional} change (a \emph{strain}) in the proper distance between freely-falling reference masses initially at rest (in this case, the LIGO mirrors). Since the strain produced at Earth by the strongest expected GW sources is of order $10^{-21}$, long interferometer arms (for LIGO, 4 km), high laser power, and extraordinary efforts to avoid any forces on the test masses are required to reach sufficient sensitivity.

\begin{figure}
\includegraphics[width=0.7\columnwidth]{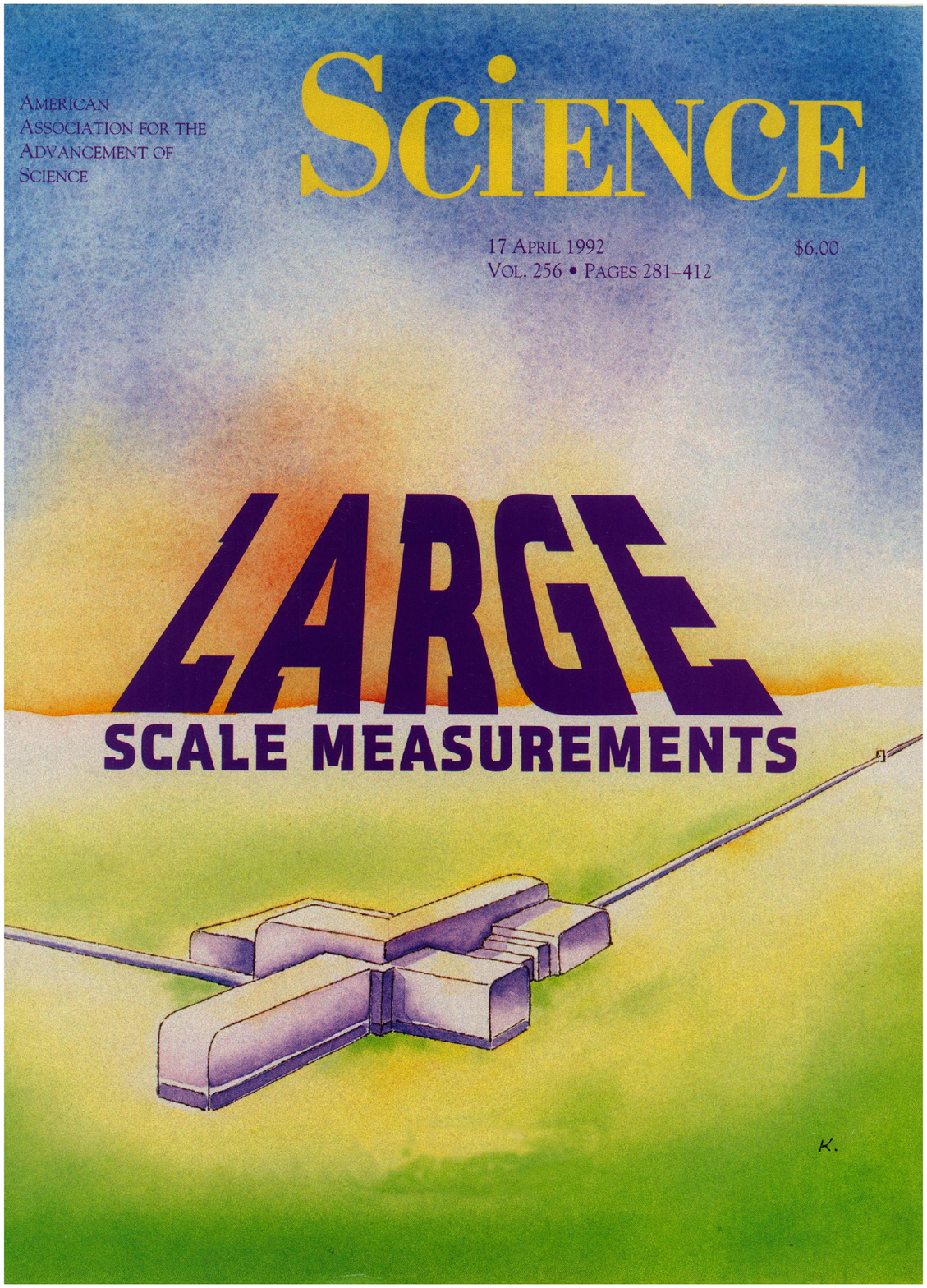} \\
\caption{Big science requires large experiments: front cover of the \emph{Science} issue including a description of the funded LIGO project \cite{Abramovici325}.}
\label{fig:ligo}
\end{figure}

In 1991 Congress approved first-year funding for LIGO. In 1992, a detailed description of the project was published in \emph{Science} \cite{Abramovici325}, and LIGO's sheer size was featured on the front cover (Figure~\ref{fig:ligo}). The Virgo project was approved in 1993 by the French Centre National de la Recherche Scientifique (CNRS), and in 1994 by the Italian Istituto Nazionale di Fisica Nucleare (INFN). 

The prospective GW sources for Earth-based interferometric detectors include compact binary coalescences (CBCs): pairs of black holes and neutron stars (in any combination). As we shall discuss in Section~\ref{sec:data-analysis}, searches based on matched filtering allow the identification of signals buried well below noise level, if their shape is known precisely.
While post-Newtonian calculations~\cite{2014LRR....17....2B} can handle the early and intermediate phases of binary inspirals, and the particle-physics-inspired \emph{effective-one-body} formalism~\cite{1999PhRvD..59h4006B} can provide sufficient precision for the late inspiral and plunge, fully non-linear numerical simulations of the Einstein equations are required to model the final merger, and to provide initial conditions for the perturbative relaxation of the final remnant black hole \cite{Berti:2009kk}. (Indeed, the \emph{quasinormal modes} of black-hole \emph{ringdown} can be handled analytically, but require numerical relativity to be connected to the merger phase.)
The need for accurate gravitational waveforms lent importance (and some urgency) to the task of modeling black-hole and neutron-star collisions, as well as the resulting GW emission, within the field of numerical relativity. 

Predicted CBC rates for first-generation interferometric detectors had uncertainties of orders of magnitude, extending to very low values \cite{Abadie:2010cf}, so it was not surprising when no detection were reported. In fact, the discussion of LIGO in Ref.~\cite{Abramovici325} had anticipated:
\begin{displayquote}
{\it
This first detector system may discover GWs. If not, experimenters will press forward with detector improvements (...), leading towards LIGO's advanced detector goals. These improvements are expected to lead to the detection of waves from many sources each year. 
}
\end{displayquote}

Starting around year 2011, LIGO and Virgo underwent extensive upgrades from their initial configurations to more advanced designs, with the goal of increasing sensitivities by an order of magnitude. This translates into thousand-fold increase in the sensitivity volume, and thus in the rate of accessible events. 
Advanced LIGO's first observing run (``O1'') took place between September 2015 and January 2016, with further runs planned starting in fall 2016 at progressively higher sensitivities, until the limits of the design are reached. As of this writing, Advanced VIRGO is expected to start taking data at the end of 2016.
 
\section{The analysis of gravitational-wave detector data} \label{sec:data-analysis}

From the standpoint of data analysis, GWs of different nature are searched for using different, customized techniques; we outline some of these efforts below.

\subsection{Burst-GW Searches}

For short, unmodeled GW transients (in GW lingo, \emph{bursts}), such as the waves from core-collapse supernovae~\cite{2011LRR....14....1F}, one first recasts the data in a time--frequency representation appropriate to describe the local energy content of the signal.
Such representations find their origin in the \emph{short-time Fourier transform} introduced by Gabor in 1946; see~\cite{Cohen1989} and~\cite{Sejdic2009} for reviews of the steady developments in this field. In the time--frequency plane, a sufficiently loud GW burst appears as a localized cluster of \emph{excess power} on top of a diffuse background of stationary detector noise. 
Search algorithms identify promising clusters, normalizing amplitudes by the baseline detector noise at different frequencies, and eliminating non-stationary detector \emph{glitches} by way of a \emph{coherence} constraint among multiple detectors.

The coherence constraint is formulated in the context of signal reconstruction. 
Specifically, the GW perturbation tensor introduced in Eq.\ \eqref{eq:gw-waves} can be rewritten more generally as
\begin{equation}
h_{ij}(t - k^l x_l) = \mathsf{e}^+_{ij}(k^l) h_+(ct - k^l x_l) + 
\mathsf{e}^\times_{ij}(k^l) h_\times(ct - k^l x_l),
\end{equation}
where $\mathsf{e}^{+}_{ij}$ and $\mathsf{e}^{\times}_{ij}$ are the symmetric, traceless, and transverse ($\mathsf{e}^{+}_{ij} k^j = \mathsf{e}^{\times}_{ij} k^j = 0$) GW polarization tensors, and $k^l$ describes the direction of propagation of the GW signal (or equivalently the sky position of its source). The signal registered by detector $A$ then becomes
\begin{equation}
\label{eq:response}
\begin{aligned}
h_A(t) &= \mathsf{d}_A^{ij} h_{ij}(t - k_l x_A^l) \\ &= F^+_A(k_l) h_+(ct - k_l x_A^l) + F^\times_A(k_l) h_\times(ct - k_l x_A^l),
\end{aligned}
\end{equation}
where $x_A^l$ describes the position of the detector and the tensor $\mathsf{d}_A^{ij}$ the orientation of its sensitive axes. The projection coefficients $F^+_A(k_l)$ and $F^\times_A(k_l)$ are known as \emph{antenna patterns}. The phases of the waveform components $h_+$ and $h_\times$ are also related  under the assumption that the signal is linearly, circularly, or elliptically polarized.
Given the data recorded by multiple detectors around a ``bright'' excess-power cluster, Eq.\ \eqref{eq:response} sets up the inverse problem for bursts, providing the basis for estimating the waveforms $h_+$ and $h_\times$, the GW polarization, and the sky location of the source (see, e.g.,~\cite{2016PhRvD..93d2004K}). Crucially, this analysis yields a measure of signal coherence, which is used to identify strong detection candidates.

In practice, these principles are implemented in sophisticated software pipelines that emphasize robustness, \emph{online} use (i.e., the processing of detector data as they are collected), low latency, and source-parameter estimation (especially of sky location, which is crucial to the electromagnetic follow up of GW detections). One such burst pipeline first reported GW150914 as a very significant transient just three minutes after the GW passed the Earth; a second pipeline confirmed the event after a few hours. Many details and further references can be found in the GW150914 burst-search companion paper~\cite{2016arXiv160203843T}. We will discuss at the end of Sec.\ \ref{sec:modeledgw} the assessment of statistical significance of GW candidates; in the meantime, let us continue our brief review of detection methods.

\subsection{Stochastic-GW Searches}

\emph{Stochastic} GW sources include the fossil radiation from the early universe~\cite{Maggiore2000} and the incoherent superposition of individually undetectable waves from compact-binary inspirals~\cite{Regimbau2011}. The random character of this radiation makes it very different than short bursts or frequency-coherent chirps from binary inspirals: while a stochastic signal will be recorded by a single detector, it will be indistinguishable from that instrument's intrinsic noise. However, thanks to the (almost) omnidirectional sensitivity of GW detectors, closely related stochastic signals will be recorded by all instruments targeting the same frequency band.
Thus, the favored search relies on \emph{cross-correlating} the data of one or more detector pairs, under the assumption that all other noise sources remain uncorrelated. This technique was used as early as 1975 for resonant-bar detectors~\cite{Hough1975}, and it was given its current mathematical formalism in~\cite{PhysRevD.59.102001}, building on well-established notions in signal processing and probability theory, such as spectral estimation and Neyman--Pearson optimal detection (see~\cite{Scharf1991} for a textbook treatment).

In the simplest case of coincident, coaligned, and identical detectors, one may form the product of the two detector outputs $s_1(t)$ and $s_2(t)$,

\begin{equation}
\label{eq:correlation}
S = \int_{-T/2}^{T/2} s_1(t) s_2(t) \,\mathrm{d}t,
\end{equation}
where $T$ is the duration of the observation, which we center arbitrarily at $t = 0$. The crucial figure of merit is the \emph{signal-to-noise} ratio $\mu/\sigma$, where $\mu$ is the expectation value of $S$ and $\sigma$ is the square root of its variance. Because the intrinsic noise of the two detectors is uncorrelated, $\mu$ depends only on the GWs, and it integrates to $T \sigma_h^2$, with $\sigma_h^2$ the total variance of the GW as registered in each instrument.\footnote{For interferometric detectors, $\sigma_h^2$ is $3H_0^2/(20 \pi^2) \int_0^\infty |f|^{-3} \Omega_\mathrm{gw}(|f|) \, \mathrm{d}f$, with $H_0$ the Hubble constant, and $\Omega_\mathrm{gw}(f) \equiv 1/\rho_\mathrm{c} \mathrm{d}\rho_\mathrm{gw}/\mathrm{d} \log f$ the logarithmic density of the GW energy that (isotropically) bathes the Universe, in units of critical energy density~\cite{PhysRevD.59.102001}.} By contrast, $\sigma$ is dominated by intrinsic detector noise; since it is the product of two uncorrelated random processes, it scales as a random walk: $\sigma^2 \propto T$ times the integrated product of the two detectors' noise spectral densities. 
The larger the $\mu/\sigma$, the more statistically confident we are that a stochastic GW signal is actually present in the data. Since $\mu/\sigma$ scales with $\sqrt{T}$, statistical significance is accumulated (slowly) with longer and longer experiments.

Of course, in actual practice things are more complicated. Most important, the product $S$, and therefore the sensitivity of a search, are reduced for non-coincident and non-aligned detector pairs. Because of the time delay between them, detectors respond to instantaneously different (if partially correlated) wavefronts; because of the difference in orientation, they register different combinations of polarizations. Thus, the GW strains observed in any two instruments will overlap only partially.
The reduction in sensitivity is quantified by the \emph{overlap reduction function} $\gamma(f)$~\cite{PhysRevD.48.2389}. The function $\gamma(f)$ is unity at $f=0$ if the detectors are aligned; it crosses zero slightly above $f = c/(2 d)$, with $d$ the distance between the detectors; and it exhibits a rapidly damped oscillation about zero for higher frequencies. The \emph{optimal filter} $S'$ that maximizes $\mu/\sigma$ is in fact given by a variant of Eq.\ \eqref{eq:correlation} that takes into account $\gamma(f)$ as well as the spectra of the GW stochastic signal and of detector noise~\cite{PhysRevD.59.102001}.

\subsection{Modeled-GW Searches}
\label{sec:modeledgw}

For GW transients whose exact shape may be known \emph{a priori} from theory (in the case of GW150914, the waves from the inspiral, merger, and ringdown of a black-hole binary), the reference search involves \emph{matched filtering}, a technique originating in radar applications in the middle of the 20th century~\cite{WainsteinZubakov1962,Turin1960}.

The need for accurate waveforms in matched filtering has been a driving motivation behind the quest for analytical and numerical solutions of GR, so it is worth understanding the basic principles of matched filtering.

In matched filtering, one looks for a known signal $h(t)$ embedded in additive noise [so the data can be represented as $s(t) = h(t) + n(t)$] by computing the signal-to-template correlation integral
\begin{equation}
\label{eq:matchedfilter}
\rho = \int_0^T \!\! s(t) \hat{h}(t) \, \mathrm{d}t = \int_0^T \!\!  h(t) \hat{h}(t) \, \mathrm{d}t \, + \int_0^T \!\! n(t) \hat{h}(t) \, \mathrm{d}t;
\end{equation}
here $\hat{h}(t)$ is the \emph{matched filter}, which is obtained by applying a noise-weighting linear operator to $h(t)$ (more about this below). For the purpose of this illustration, we can take $\hat{h}(t) = h(t)$, which is appropriate for \emph{white noise}.\footnote{White noise is a random signal with constant power spectral density. In applications that involve discretely sampled data, it is useful to think of white noise as a sequence of uncorrelated random variables with zero mean and finite variance.}
Again, the ratio $\mu/\sigma$ (in this case, for the variable $\rho$) is a measure of statistical confidence that the signal $h(t)$ is present in the data. The two integrals on the right-hand side of Eq.\ \eqref{eq:matchedfilter} correspond to $\mu$ and $\sigma$ respectively: the first accumulates as $\sim h_0^2 T$, where $h_0$ is the characteristic amplitude of the signal $h(t)$; the second grows only as $(\tau_0 T)^{1/2} n_0 h_0$ (in random-walk fashion), where $n_0$ is the characteristic amplitude of the noise, and $\tau_0$ is related to the noise timescale (e.g., for band-limited white noise $\tau_0 \sim 1/f_\mathrm{max}$). Since $\mu/\sigma \sim (T/\tau_0)^{1/2} h_0/n_0$, we see that the matched filter allows detection even if $h_0 / n_0$ is significantly less than one. The enhancement is again proportional to $\sqrt{T}$.

More formally, the random variable $\rho$ is used as a \emph{detection statistic}~\cite{Scharf1991} as follows: in the presence of noise alone, the probability distribution $p_n(\rho)$ is dictated by the properties of noise; in the presence of noise and signal, the probability distribution $p_{n+h}(\rho)$ becomes displaced to larger values. If we measure a sufficiently large $\rho$, we can conclude with high confidence that the signal $h$ is present in the data. To quantify that confidence, we set a \emph{false-alarm probability} $P_\mathrm{FA}$ and obtain the \emph{threshold} $\rho_\mathrm{tr}$ such that the cumulative probability $P_n(\rho > \rho_\mathrm{tr}) = P_\mathrm{FA}$. In other words, if we claim a detection whenever $\rho > \rho_\mathrm{tr}$, we are only wrong $P_\mathrm{FA}$ of the times. Conversely, this scheme results in falsely dismissing a fraction $P_{n+h}(\rho < \rho_\mathrm{tr}) \equiv P_\mathrm{FD}$ of true signals. The matched filter is the \emph{optimal} linear filter in the sense that it minimizes $P_\mathrm{FD}$ for a fixed $P_\mathrm{FA}$.

The simplified picture that we have just drawn describes a search for a signal of known shape, occurring at a known time. The generalization to unknown source parameters involves trying out many different signal shapes. For some parameters, the variation of the signal can be handled with analytical techniques; such is the case of overall amplitude, initial phase, and merger time of inspiral signals, all of which map into simple transformations of the filter $\hat{h}$ in the Fourier domain (see, e.g.,~\cite{Maggiore2008}). To tackle the variation due to other parameters, such as the component masses in a binary inspiral, it is necessary to evaluate the detection statistic across a \emph{bank} of \emph{signal templates}. The templates must be placed strategically across the space of source parameters: densely enough that no real signal is missed because its phase evolution matches no template in the bank; sparsely enough that the computation remains feasible. This is an interesting geometrical problem that can be formulated in terms of differential geometry~\cite{1996PhRvD..53.6749O}, \emph{ad hoc} tilings~\cite{2003PhRvD..67j2003A}, 	
as well as lattice-based and randomized sphere coverings~\cite{2007CQGra..24S.481P,2009PhRvD..79j4017M}. If we filter detector data against a template bank, we need to raise the detection threshold to take into account the fact that we are running multiple independent ``experiments,'' each of which can result in a false alarm.\footnote{The actual number of independent trials is difficult to control analytically because of partial template correlations, and is best determined by experiment.}
For the Advanced LIGO search that led to the detection of GW150914 as a modeled transient, 250,000  templates were placed across parameter space, with each component mass ranging from 1 to 100 solar masses~\cite{2016arXiv160203839T}.

The other immediate generalization is to colored noise. Under the restrictive but enabling assumption that the instrument noise $n(t)$ is Gaussian and stationary, its \emph{sampling distribution} is determined entirely by its correlation function, or equivalently by its \emph{power spectral density} $P(f)$ (see, e.g.,~\cite{Maggiore2000}). To wit,
\begin{equation}
\label{eq:noiseproduct}
p_n(n) \propto \exp\left\{ -2 \int_0^\infty \frac{\tilde{n}^*(f) \tilde{n}(f)}{P(f)} \, \mathrm{d}f \right\} =
\exp\left\{-\langle n|n \rangle/2\right\},
\end{equation}
where $\tilde{n}(f)$ is the Fourier transform of $n(t)$, $\tilde{n}^*(f)$ is its complex conjugate, and $P(f)$ is defined by the ensemble (i.e., noise-realization) average $\langle \tilde{n}^*(f) \tilde{n}(f')\rangle = \delta(f - f') P(|f|) / 2$. The power spectral density can be estimated empirically from a nearby stretch of data, under the assumption that it is not affected significantly by the presence of GW (see~\cite{PhysRevD.85.122006} for the details of a sophisticated implementation). It is straightforward to show that in the presence of colored Gaussian noise the optimal matched filter can be written in the frequency domain as $\hat{h}(f) = \tilde{h}(f) / P(f)$~\cite{WainsteinZubakov1962}.

In fact, while the assumption that noise is stationary and Gaussian is pervasive in the theoretical development of GW data-analysis methods, it cannot be trusted to establish the \emph{true} false-alarm probability of detection candidates, since interferometric detectors are neither stationary nor Gaussian. Instead, the \emph{background} rate of noise-induced false alarms is determined empirically.  
For instance, in searches that require coincident candidates in multiple detectors, the background is measured by artificially sliding the time axis of the data in one detector, and then performing the coincidence analysis. Any resulting joint candidates are necessarily the product of noise alone. This very procedure established that the false-alarm probability of GW150914 was (much) less than one in 203,000 years~\cite{Abbott:2016blz}.

\subsection{Parameter Estimation}

Neverthless, Eq.\ \eqref{eq:noiseproduct} is the basis of the probabilistic treatment of detection and parameter estimation for GW signals of known shape; this is with good reason, since detected signals are likely to be found in well-behaved noise; furthermore, the assumption of Gaussian noise can be validated \textit{a posteriori}.

Together with the assumption of additive noise (again, that the detector data $s(t)$ equals GW signal $h(t)$ plus instrument noise $n(t)$), Eq.\ \eqref{eq:noiseproduct} leads to the \emph{likelihood} $p(s|h(\theta))$ that the data contain a GW signal determined by the parameter vector $\theta$:
\begin{equation}
p(s|h(\theta)) \propto \exp\left\{-\bigl\langle s-h(\theta)\big|s-h(\theta) \bigr\rangle/2\right\},
\end{equation}
where the product $\langle \cdot | \cdot \rangle$ has the same form as in Eq.\ \eqref{eq:noiseproduct}.
Among the templates $\{h(\theta_i)\}$ in a bank, the template that results in the largest $\rho$ [of Eq.\ \eqref{eq:matchedfilter}] is also the template that maximizes the likelihood $p(s|h(\theta))$. If we extend the bank to a continuous family over parameter space, we obtain the \emph{maximum-likelihood} point estimate of the source parameters as~\cite{Maggiore2000}
\begin{equation}
\theta^\mathrm{ML} = \mathrm{maxloc}_\theta \, p(s|h(\theta)).
\end{equation}
The error of this estimate is quantified by the Fisher information matrix $F_{\mu \nu} = \langle \partial_\mu h | \partial_\nu h \rangle|_{\theta^\mathrm{ML}}$, where the $\partial_\mu$ denote partial differentiation with respect to the source parameters~\cite{2008PhRvD..77d2001V}. Specifically, for sufficiently strong signals the error vector $\theta^\mathrm{ML} - \theta^\mathrm{true}$ is distributed normally with parameter covariance given by $F^{-1}_{\mu \nu}$. (For weaker signals detected near the threshold, the story is more complicated~\cite{2008PhRvD..77d2001V,2011PhRvL.107s1104V}.) The probability distribution of the errors refers to a hypothetical infinite sequence of experiments (each with a different noise realization) that result in the same $\theta^\mathrm{ML}$, a construct typical of frequentist (a.k.a., classical) statistics.

In fact, the GW data-analysis community has by now largely transitioned to Bayesian methods for parameter estimation~\cite{Gregory2005}. This happened for several reasons: to distill maximum information out of the data we will have from rare (at least initially) detections; to incorporate prior information from astrophysical theory or non-GW observations; and to draw physical conclusions jointly from multiple sources (e.g., about astronomical populations~\cite{2016arXiv160203842A}, or violations of general-relativistic predictions~\cite{2016arXiv160203841T}).
In this respect, GW analysts have anticipated a broader movement in the larger astronomical community~\cite{2012arXiv1208.3036L}.

In Bayesian inference~\cite{Gregory2005}, one thinks of a GW observation in the data $s(t)$ as updating the \emph{prior} probability density $p(\theta)$ of the source parameters into their \emph{posterior} density, by way of Bayes' theorem:\footnote{Mathematically, Bayes' theorem is nothing more than a restatement of the law of compound probabilities, but the name carries the import of its interpretation in terms of rational (and axiomatized) degree of belief~\cite{Jaynes2003}.}
\begin{equation}
\label{eq:bayes}
p(\theta|s) = \frac{p\bigl(s|h(\theta)\bigr) p(\theta)}{\int p\bigl(s|h(\theta)\bigr) p(\theta) \, \mathrm{d}\theta}.
\end{equation}
The Bayesian view is that the posterior, taken as a whole, quantifies our informed belief about the signal. It is still possible to distill the posterior into point estimators, such as conditional means, and simple measures of uncertainty, such as parameter covariances. The denominator in Eq.\ \eqref{eq:bayes} is the \emph{Bayesian evidence} $P(s)$, a measure of the overall credibility of the data as having been generated according to the mathematical model encoded in the prior and likelihood. A Bayesian analyst may compare the evidence of different models to decide between them: for instance, between template families representing black-hole vs.\ neutron-star binaries, or even between a signal model vs.\ a noise-only model, which sets up a Bayesian detection scheme~\cite{Scharf1991}.

In most cases, the posterior must be explored and integrated numerically. This is a difficult problem: because of the moderately high dimensionality of parameter space, all schemes that involve a structured walkthrough of parameter space (e.g., along multidimensional grids) require huge computational resources. The breakthrough in Bayesian computation took place in the 1990s, when statisticians began to apply \emph{Markov Chain Monte Carlo} to inference (see, e.g.,~\cite{Gilks1996} for an introduction). The technique originates with the seminal paper by Metropolis, the Rosenbluths, and the Tellers~\cite{Metropolis1953}: the idea is to explore statistical ensembles in stochastic fashion on a computer, (crucially) sampling states directly from the target distribution:\footnote{The reason this is so important is that, for most distributions of interest, a state selected randomly across parameter space is likely to have very small probability, thus contributing very little to statistical integrals and distributions.}
\begin{displayquote}
{\it
Instead of choosing configurations randomly, then weighting them with $\exp(-E/kT)$, we choose configurations with a probability $\exp(-E/kT)$ and weight them evenly.~\cite{Metropolis1953}
}
\end{displayquote}
This becomes possible thanks to a clever scheme (the \emph{Metropolis rule}, later generalized in many ways and directions~\cite{Gilks1996}) to conditionally accept or reject each step of a random walk, in such a way that the trajectory approaches asymptotically the distribution of interest---whether $\exp(-E/kT)$ or the Bayesian posterior. Markov Chain Monte Carlo was first applied to GW searches by Christensen and Meyer in 1998~\cite{1998PhRvD..58h2001C}. Its current implementation in the LIGO software library~\cite{2015PhRvD..91d2003V} relies on GW-signal-specific customizations of a few broadly applicable modern variants of Monte Carlo, such as parallel tempering~\cite{Geyer1991} and nested sampling~\cite{Skilling2006}.
(See instead~\cite{Green2015} for a broader review of the state of the art in Bayesian computation.) In the case of GW150914, these techniques were able to confidently determine the masses of the component black holes, the distance and sky location of the system, and much more~\cite{2016arXiv160203840T}; and even to test key predictions of GR for such a binary~\cite{2016arXiv160203841T}.

\section{Numerical Relativity in the 21st Century}

Driven to a large extent by the search for direct detection of GWs, a Binary Black Hole Grand Challenge Alliance (GCA) was funded by the NSF from 1993 to 1998, involving $40$ researchers at $10$ institutions. Its goal was to develop mathematical and numerical techniques, along with a high performance computing (HPC) infrastructure, specifically tailored to the binary black hole problem that would allow stable simulations of generic collisions of binary black holes. In particular, to move from head-on collisions to the more general case of black holes which inspiral around each other while emitting GWs, their orbits shrinking, and eventually merging to form a new black hole which, assuming non-linear stability of black holes under large perturbations (Section~\ref{sec:bh-stability}), it should decay to a member of the Kerr solution according to the no hair theorem (Section~\ref{sec:no-hair}). 

The GCA started a number of efforts which continued well beyond the funding of the project, regarding manifestly hyperbolic formulations of the Einstein equations, methods for extracting gravitational radiation from numerical space-times, evolution approaches ranging from Cauchy to characteristic formulations, HPC tools for large scale parallel computations and visualization, and adaptive numerical techniques. The latter was pioneered by Matt Choptuik, in his studies of gravitational collapse which led to his discovery of critical behavior~\cite{Choptuik:1992jv}.

The first attempts at black hole excision were made, whereby one uses the fact that the region inside a black hole is causally disconnected from the outside to excise it from the computational domain. Practical approaches to outer boundary conditions were also explored, although the mathematical formulation as an initial-boundary-value problems, and its well posedness, was not carefully addressed until the 2000's (Section~\ref{sec:ibvp}). From a physical point of view, the head on collision of Misner's initial data was revisited in great detail, though still with limited initial separations. 
A 1998 summary of the GCA by its PI, Richard Matzner, can be found in~\cite{CGA-MoG}. 

After the GCA, or by the end of it, a systematic mathematical analysis of the Einstein equations as an initial-boundary value problem was initiated (earlier in Section~\ref{sec:ibvp} we discussed some of this work, including~\cite{Friedrich99,Kreiss:2006mi,Kreiss:2008ig}). 
At the same time, a thorough numerical analysis was being done to develop more accurate and reliable numerical methods for the Einstein equations.
Both activities involved an increasing level of interaction between numerical relativists from the physics community and specialists in numerical analysis and PDEs from the mathematics community.
This interaction happened by design rather than accident, due to the forsight of a number of people working in these three areas.
Starting around 2002, a number of workshops were held with the specific goal of bringing together researchers from all three communities.
These began with a {\it Hot Topics Workshop} on numerical relativity held at the IMA in June 2002, followed by a year-long Mathematical/Numerical Relativity Visitor Program at Caltech from 2002-2003 (and continuing for some time into 2003-2004).
These led to additional workshops focussing on both mathematical and numerical relativity organized by some of the participants of these first two meetings.
These included an AIM/Stanford Relativity workshop in 2003, an IPAM/UCLA {\it Geometric Flows} workshop in 2003, a second IMA Workshop in 2004, a BANFF/BIRS Numerical Relativity Workshop in 2005, and a Mathematical Relativity Workshop Visitor Program at the Isaac Newton Institute at Cambridge in 2005.
Minisymposia were also organized at larger mathematics meetings that  
involved speakers from both the mathematics and numerical relativity communities, including sessions at the Miami Waves Workshop in 2004.
(Various subsets of the authors of this article attended all of these meetings, and met for the first time at one of the meetings.)

These interactions helped accelerate the infusion of useful ideas and tools from the mathematics community into numerical relativity.
For example, for some time it had been thought that the simulation of black holes would benefit from advanced high resolution shock capturing methods, even in the absence of matter fields. 
However, it was pointed at the 2002 IMA Workshop by participants that in the vacuum case, the Einstein equations are linearly degenerate, and shocks are not expected.
That is, while the solution is finite, it is expected to be smooth.
Therefore, on the contrary, one expects high order or spectral methods to be best suited for such cases, and indeed, they became dominant in the 2000's (for example~\cite{specweb}). 
High resolution simulations of single black holes using spectral methods showed that instabilities present in some simulations would not be ruled out by more powerful supercomputers, but a deeper understanding of the continuum properties of the field equations was needed.
During the intensive interactions between researchers during this period, a number of known numerical analysis techniques were refined and tuned for the Einstein equations, and some completely new techniques were developed to handle the unique problems that arise with numerical simulation of the Einstein equations.
By 2005, what had seemed in 2002 to be an almost intractible simulation problem was starting to crack.

In 2005, at the Banff International Research Station in Canada {\it Numerical Relativity} meeting, Pretorius reported the first set of long-term simulations of binary black holes inspiraling around each other, merging and decaying to a stationary black hole, with no signs of growing instabilities and with the accuracy to extract waveforms from the numerical space-time~\cite{Pretorius:2005gq}. One of the key insights in this  breakthrough was a previously proposed constraint damping mechanism~\cite{Brodbeck:1998az,Gundlach:2005eh}, originally stemming from computational fluid dynamics and quoted as privately proposed to relativists by Heinz-Otto Kreiss in Ref.~\cite{Brodbeck:1998az}.

Soon thereafter, two groups reported at the same meeting in NASA-Goddard the independent discovery of a very different approach to produce equally stable and accurate long-term simulations based on the BSSN formulation of the Einstein equations~\cite{Baker:2005vv,Campanelli:2005dd}. This approach has since then been coined  {\it moving punctures} because the key ingredient is to use coordinate conditions which allow the black hole punctures to move across the computational domain.  Punctures do not represent the location of black holes; instead, they represent a compactification of each ``infinity'' region in non-trivial topological constructions of multiple black holes, see for example~\cite{Brown:2007tb}.

\begin{figure}
\mbox{\includegraphics[width=0.65\columnwidth,angle=270]{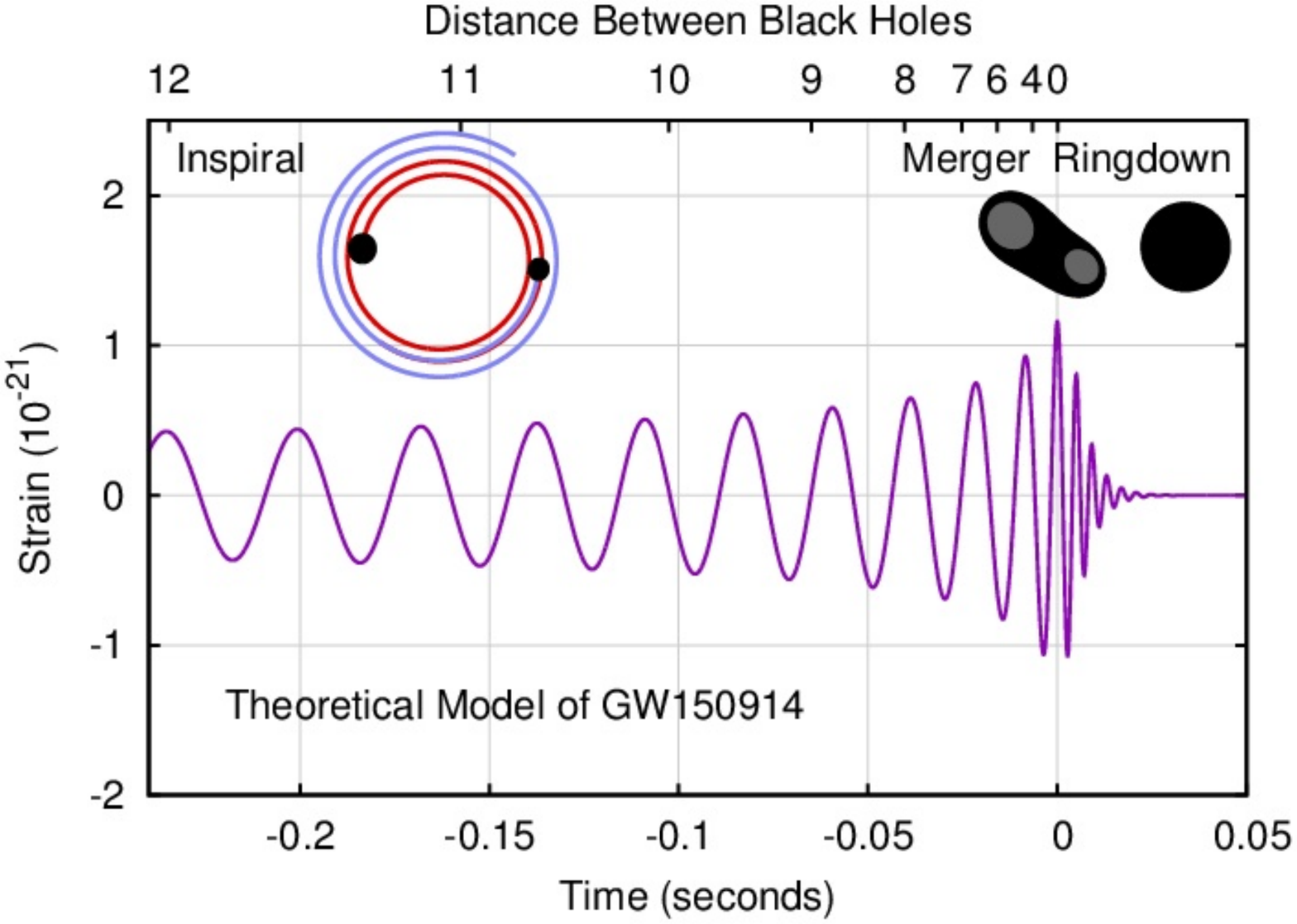}}
\hspace*{-0.6cm}
\mbox{\includegraphics[width=0.88\columnwidth]{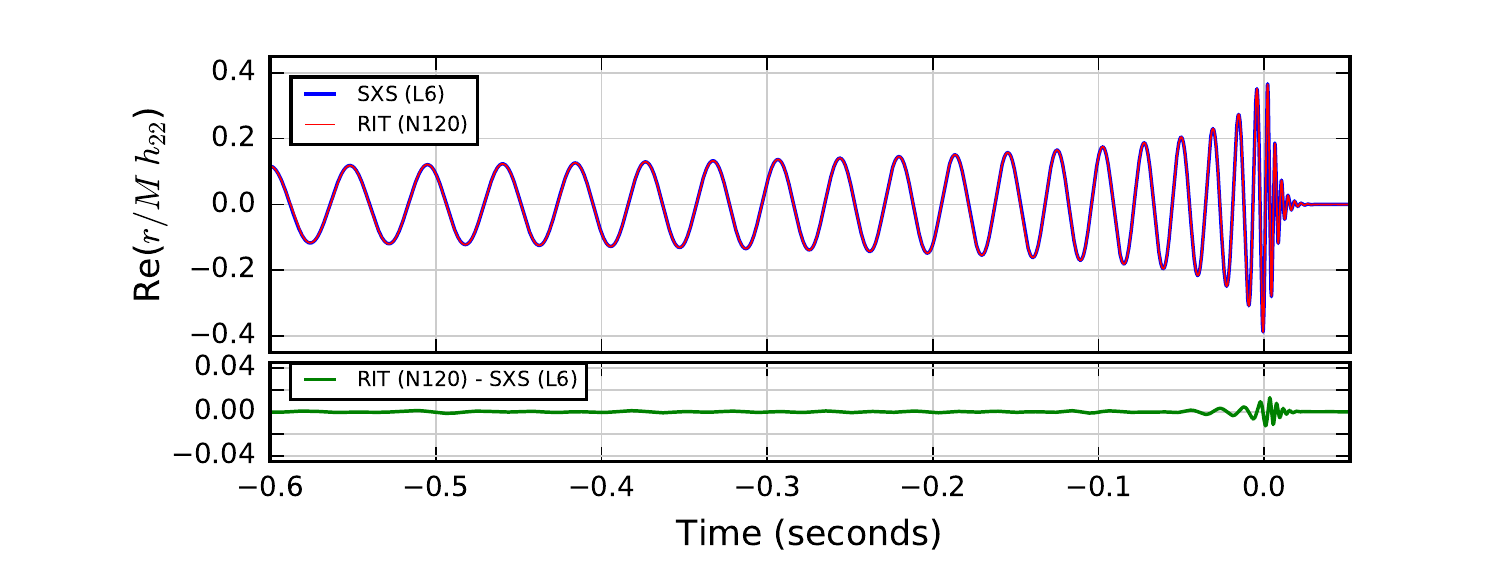}}
\caption{
Illustration of the GW150914 detection.
Top image shows the inspiral, merger, and ringdown phases of the GW150914 event relative to the dominant part of a theoretical model of the emitted gravitational waveform.
Bottom image shows two numerical simulations of a binary black hole system producing waveforms closely matching GW150914.
The two simulations were produced by different computer codes, developed by separate groups, using different numerical techniques.
Top image courtesy of the Center for Computational Relativity and Gravitation at the Rochester Institute of Technology~\cite{RIT-Website}.
Bottom image courtesy of Geoffrey Lovelace in the Gravitational Wave Physics and Astronomy Center at the California State University, Fullerton~\cite{CSUF-Website}.
}
\label{fig:bbh}
\end{figure}

In 1987~\cite{hawking1989three} Thorne had predicted that
\begin{displayquote}
{\it
For black hole [collisions] numerical relativity seems likely to give us, within the next five years, a detailed and highly reliable picture of the final coalescence and the wave forms it produces, including the dependence on the hole's masses and angular momenta. Comparison of the predicted wave forms and the observed ones will constitute the strongest test ever of general relativity.
}
\end{displayquote}
It actually took about two more decades; that is, around four more decades after the original simulations of Hahn and Lindquist, to get to the point in which any single configuration could be systematically simulated. The long quest to be able to simulate binary black holes came just in time for the new era of GW science. In particular, numerical simulations have been critical in developing semi-analytical and phenomenological models of the GW emission from CBCs. A recent review of numerical relativity as a probe of gravity in its strong regime, part of a book commemorating the 100th anniversary of GR, can be found in~\cite{Choptuik:2015mma}.

\section{The Discovery of GW150914} \label{sec:gw-detection}

The approximately hour-long press conference that we mentioned in Section~\ref{sec:ligo-announcement} included NSF director France C\'ordova, LIGO Laboratory Executive Director David Reitze, LIGO Scientific Collaboration spokesperson Gabriela Gonzalez, and LIGO co-founders Rainer Weiss and Kip Thorne (Ronald Drever could not attend for health reasons). 
The announcement provided details about the first direct detection of GWs from a binary black-hole coalescence in data collected by the two Advanced LIGO detectors on September 14, 2015, shortly after the detectors were turned on. The NSF press release is available at~\cite{NSF-LIGO-press}.

The event was named GW150914. Analysis of the data showed that the waves were generated by the coalescence of two black holes, with masses $\simeq 29$ and $36$ times that of our Sun, inspiraling and then merging to form a final Kerr black hole of $\simeq 62$ solar masses. The measured signal was rather short ($0.2$ s), covering the final $8$ cycles of the coalescence. The detection was unambiguous, with a \emph{measurable} significance greater than $5 \sigma$, for a false alarm rate of less than one event per $203,000$ years. (This was an upper limit, corresponding to the empirical background that could be explored using time slides. The actual significance was arguably much higher). It was reported by the LSC collaboration that 50 million CPU core hours (on recent high-end ``conventional'' cores) were devoted to the data analysis of the signal. The production discovery analyses were run on dedicated LIGO Data Grid clusters, XSEDE and the Open Science Grid (in particular on {\it Comet}, at the San Diego Supercomputer Center, and {\it Stampede}, at the Texas Advanced Computing Center).

The signals recorded at the two LIGO sites showed exquisite agreement, and were observed with a combined signal-to-noise ratio of $24$. The time delay between the sites was $7$ milliseconds, and the event released the energy equivalent of three solar masses into GWs, with a peak luminosity $\simeq 3 \times 10^{56}$ erg/s, greater than the visible light emitted (continuously) by the rest of the Universe. As Kip Thorne commented:
\begin{displayquote}
{\it
It is by far the most powerful explosion humans have ever detected except for the Big Bang.
}
\end{displayquote}

The LIGO and Virgo collaborations had agreed that a detection would be presented publicly only after peer review, and indeed a letter describing GW150914 appeared in the Physical Review on February 11 \cite{Abbott:2016blz}.
The excitement was such that as soon as the press conference began, the Physical Review Letters webpage was overwhelmed by more than 10,000 downloads per minute, to the point that it remained inaccessible until the journal could add more servers.
Several companion papers, covering many complementary aspects of the detection and of its theoretical consequences, were also released on that day (see \url{papers.ligo.org} for links). Issue 8 of the \emph{LIGO Magazine} \cite{LIGO-mag8} presents a collection of articles about LIGO, the discovery of GW150914, its implications, and the future of GW astronomy. 

On June 15, 2016, the LIGO and Virgo collaborations announced a second highly statistically significant detection of GWs from the coalescence of two black holes~\cite{Abbott:2016nmj}. This signal, named GW151226, was recorded by the two LIGO detectors on December 26, 2015, toward the end of the first Advanced LIGO observation run (``O1'').
The black holes responsible for the signal are lighter than for GW150914, with masses $\simeq 14$ and $8$ times that of the Sun (although uncertainties are large) 
The signal was weaker than GW150914 but it lasted longer (1 s and 55 cycles) and it ended at a higher frequency within the sensitive detector band. Again the measured significance was $5 \sigma$, and the true significance arguably much higher. LIGO plans to start a second observation run (O2) of six months with slightly higher sensitivity in fall 2016, to be joined by VIRGO soon thereafter. With increasing sensitivities and longer runs, the expectation is that detections will become almost routine (up to tens per year), and the hope that black-hole binaries will be joined in the ``catch'' by other systems, especially those involving neutron stars, which could have observable electromagnetic counterparts.

\section{Opportunities for further mathematical developments}

GW science faces a number of challenges in mathematics, computational science, and data analysis, in addition to ongoing work in instrumentation, physical and astrophysical modeling, and in developing new ways to explore the Universe using GW detectors.
We mention just a few of these challenges to close this article, highlighting not only past synergies between the mathematical sciences and GW science, but also what might lay ahead. 

The most interesting challenges in the statistical theory of GW detection and parameter estimation revolve on the astrophysical and theoretical interpretation of observed GW sources.
For instance, how does one use a catalog of CBC observations to constrain populations of binaries across the Universe, or the physical processes that led to their formation? What is the best statistical formulation to test GR through CBC signals? How can we robustly take into account the systematic uncertainties due to the calibration of detectors, imperfect waveform modeling, and other perturbations?

Because extracting physical parameters from individual observed signals requires accurate theoretical waveforms as a function of the parameters, the field faces a major obstacle in the curse of dimensionality, both for the forward problem (modeling: parameters to waveforms) and inverse problem (inference: signals to parameters).
For binary black-hole waveforms, even under the simplifying assumption of negligible orbital eccentricity, modeling involves an 8-dimensional parameter spaces (two masses and two spin vectors), while inference requires exploring a 15-dimensional parameter space that includes also \emph{explicit} signal-presentation parameters such as source position, orientation, and merger time.

Now, finding solutions to the Einstein equations is numerically intensive: hundreds of thousand of hours of computing time for the final phase of inspiral--merger--ringdown signals, which cannot be modeled analytically with sufficient accuracy. 
Even with the largest supercomputers available at the time of writing, the full parameter space of interest cannot be surveyed directly, so LIGO analysts have relied on semi-analytical phenomenological waveforms calibrated to a small number of numerical simulations \cite{2016arXiv160203840T}.
Techniques based on state-of-the-art reduced order modeling for parameterized systems (see \cite{Blackman:2015pia,Canizares:2014fya} and references therein) are also being explored and commissioned; in this case one directly interpolates among a small number of numerical waveforms, without the intermediate step of developing a parameterized phenomenological representation.
Once waveforms are available, reliable statistical inference requires sophisticated sampling schemes to explore parameter space efficiently.

Both the forward and inverse problems have inherent sources of uncertainty from modeled and unmodeled sources of noise, numerical errors in simulations, approximations in initial and boundary conditions, and even the possibility of corrections to Einstein's theory of gravity. Moreover, in these theoretical studies instrumental noise is usually assumed (as in many other fields) to be Gaussian and stationary, which is almost never the case. While GW science has developed both general and \emph{ad hoc} techniques to deal with these uncertainties, useful approaches may come from the field of Uncertainty Quantification, which relies on different tools and formulations than Bayesian inference.

There are two very active, and somewhat distinct, research programs in the mathematics community that focus on developing a more complete analytical understanding of two different aspects of the Einstein equations.
The first of these programs involves primarily open problems associated with the solution theory for the evolution equations; there are still a number of important remaining issues in addition to the ones we have already mentioned in this article. 
For example, apart from proving that Kerr black holes are stable under small perturbations, one wishes to have a complete mathematical description for the gravitational collapse of matter to form a black hole.
Here, the ``holy grail'' consists in proving the \emph{cosmic censorship conjecture}, which essentially states that the Cauchy evolution of regular, asymptotically flat initial data should not produce any naked singularities (in other words, singularities may form but they must be hidden inside a black hole).
So far, due to groundbreaking work by Christodoulou, a fairly complete picture has emerged in certain spherically symmetric models including the collapse of a dust ball and the critical collapse of a massless scalar field.
However, the general case still remains widely open.
For more details and recent developments see the introduction in Christodoulou's monograph~\cite{Christodoulou09}.

The second of these research programs involves primarily open problems associated with solution theory for the Einstein constraint equations. 
One of the major activities in this program has been to complete the theory for the conformal method, and this is currently undergoing rapid development.
It was hoped that the new analysis frameworks developed in 2008 that led to the first ``far-from-CMC'' existence results would lead to a complete solution theory.
However, after it was shown in 2011~\cite{M11} that multiple solutions are possible in the non-CMC case, a number of additional techniques were developed that have led to a more refined understanding of the conformal method.
In~\cite{DGH10}, scaling and blowup techniques were developed for the conformal method, giving a new approach to obtain non-CMC existence results~\cite{GiNg14a,GiNg15a}; this was further refined in~\cite{Nguyen:2015}, giving the best characterization to date for multiplicity of general solutions in the non-CMC case.
Analytic bifurcation theory and numerical continuation methods are now also being used where possible~\cite{DW07,HoMe12a,Premoselli:2015,ChruscielGicquad:2015,DiltsHolstMaxwell15} to characterize fold and bifurcation phenomena in the conformal method.
These studies could point the way to generalizations of the conformal method, such as the \emph{drift system}~\cite{Maxwell:2014a,Maxwell:2014b,Maxwell:2014c}, that may provide better parameterizations of the initial data for GR in the truly non-CMC setting.

These are just a few examples.
Many additional synergies, including unforeseen ones, between mathematics, computational science, data analysis, and gravitational wave science, are expected to play crucial roles, just as they have over the past $100$ years. 

\vspace{12pt}

\section{Acknowledgments}

We thank the reviewers, as well as David Shoemaker, who read preliminary drafts, for their feedback and comments.  
This work was supported in part by NSF grants PHY-1500818, DMS/FRG-1262982, and DMS/CM-1217175 to the University of California at San Diego, by NSF grant PHY-1404569 to the California Institute of Technology, by CONACyT grant No. 271904, and by a CIC grant to Universidad Michoacana.
Part of this research was performed at the Jet Propulsion Laboratory, under contract with the National Aeronautics and Space Administration. 

\bibliographystyle{abbrv}
\bibliography{references,ams_refs}

\end{document}